\documentstyle[11pt,epsfig]{article}
\newcommand \be{\begin{equation}}
\newcommand \bea{\begin{eqnarray}}
\newcommand \ee{\end{equation}}
\newcommand \eea{\end{eqnarray}}
\newcommand{\lp}{\left(}
\newcommand{\rp}{\right)}

\begin{document}

\title{Finite-time singularity in the dynamics \\of the world population,
economic and financial indices. \\ 
{\normalsize Running title: Finite-time singularity in world population growth}}

\author{Anders Johansen$^1$ and Didier Sornette$^{1,2,3}$\footnote{corresponding
author}\\
$^1$ Institute of Geophysics and
Planetary Physics\\ University of California, Los Angeles, California 90095\\
$^2$ Department of Earth and Space Science\\
University of California, Los Angeles, California 90095\\
$^3$ Laboratoire de Physique de la Mati\`{e}re Condens\'{e}e\\ CNRS UMR6622 and
Universit\'{e} de Nice-Sophia Antipolis\\ B.P. 71, Parc
Valrose, 06108 Nice Cedex 2, France \\
e-mails: anders@moho.ess.ucla.edu and sornette@moho.ess.ucla.edu}

\date{\today}
\maketitle

\abstract{Contrary to common belief, both the Earth's human population and its 
economic output have grown faster than exponential, i.e., in a
super-Malthusian mode, for most of the known
history. These growth rates are compatible with a spontaneous singularity 
occuring at the {\it same} critical time $2052 \pm 10$ signaling an abrupt 
transition to a new regime. The degree of abruptness can be infered from
the fact that the maximum of the world population growth rate was reached
in $1970$, i.e., about $80$ before the predicted singular time, 
corresponding to approximately $4\%$ of the studied time interval over which
the acceleration is documented. This rounding-off of the finite-time 
singularity is probably due to a combination of well-known finite-size 
effects and friction and suggests that we have already entered the 
transition region to a new regime. 
In theoretical support, a multivariate analysis coupling 
population, capital, R\&D and technology shows that a dramatic acceleration in the 
population during most of the timespan
can occur even though the isolated dynamics do not exhibit it. 
Possible scenarios for the cross-over and the new regime are 
discussed.}

\vskip 1cm
Physica A 294 (3-4), 465-502 (15 May 2001)

\newpage

\setcounter{page}{1}

\section{Introduction}

Both the world economy as well as the human population have grown at a 
tremendous 
pace especially during the last two centuries. It is estimated that 2000 
years ago the population of the world was approximately 300 million and for 
a long time the world population did not grow significantly, since periods 
of growth were followed by periods of decline. It took more than 1600 years 
for the world population to double to 600 million and since then the growth 
has accelerated. It reached 1 billion in 1804 (204 years later), 2 billion in 
1927 (123 years later), 3 billion  in 1960 (33 years later), 4 billion in 1974 
(14 years later), 5 billion in 1987 (13 years later) and 6 billion in 1999
(12 years later). This rapidly accelerating growth has raised sincere worries 
about its sustainability as well as concerns that we humans as a result might 
cause severe and {\it irreversible} damage to eco-systems, global weather 
systems etc \cite{Cohenscience,Hern}. At, what one may say the other 
extreme, the optimists  expect that the innovative spirit of mankind will be 
able to solve the problems associated with a continuing increase in the 
growth rate \cite{vonFoerster,Simon}. Specifically,
they believe that the world economic development will continue as a successive
unfolding of revolutions, {\it e.g.}, the Internet, bio-technological and 
other yet
unknown innovations, replacing the prior agricultural, industrial, medical
and information
revolutions of the past. Irrespective of the interpretation, the important
point is the
presence
of an {\it acceleration} in the {\it growth rate}. Here, it is first shown 
that,
contrary to common belief, both the Earth human population as well as its
economic output have grown faster that exponential for most of the known 
history and most strikingly so in the last centuries. Furthermore, we will 
show that both the population growth rate and the economic growth rate
are consistent with a spontaneous singularity at the {\it same} critical
time $2052 \pm 10$ and with the same characteristic self-similar geometric
patterns (defined below as log-periodic oscillations). Multivariate dynamical
equations coupling population, capital and R\&D and technology can indeed
produce such an ``explosion'' in the population even though the isolated 
dynamics do not. In particular, this interplay provides an explanation of 
our finding of the same value of the critical time $t_c (\approx 2052 \pm 10)$
both for the population and the economic indices. As a consequence,
even the optimistic view has to be revised, since the acceleration of the
growth rate
contains endogenously its own limit in the shape of a finite-time
singularity to be
interpreted as a transition to a qualitatively new behaviour. Close to the
mathematical
singularity, finite-size effects will smoothen the transition and it is 
quite possible that Mankind may already have entered this transition phase. 
Possible scenarios for the cross-over
and the new regime are discussed.

\subsection{The logistic equation and finite-time singularities}

As a standard model of population growth, Malthus' model assumes that the size
of a population increases by a fixed proportion $\tau$ over a given period of
time independently of the size of the population and thus gives an exponential
growth. The logistic equation attempts to correct for the resulting unbounded
exponential growth by assuming a finite carrying capacity $K$ such that the
population instead evolves according to
\be
{d p \over dt} = r p(t) \left[K - p(t)\right]~. \label{ajdak}
\ee
Cohen and others (see \cite{Cohenscience} and references therein) have put
forward idealised models taking into account interaction between the human
population $p(t)$ and the corresponding carrying capacity $K(t)$ by assuming
that $K(t)$ increases with $p(t)$ due to technological progress such as the
use of tools and fire, the development of agriculture, the use of fossil fuels,
fertilisers {\it etc.} as well an expansion into new habitats and the removal
of limiting factors by the development of vaccines, pesticides, antibiotics,
{\it etc.} If $K(t) > p(t)$, then $p(t)$
explodes to infinity after a finite time creating a singularity. In this case,
the limiting factor $-p(t)$ can be dropped out and, assuming a simple power
law relationship $K \propto p^{\delta}$ with $\delta >1$, (\ref{ajdak})
becomes
\be
{d p \over dt} = r [p(t)]^{1+\delta}~, \label{aafajdak}
\ee
where the growth rate accelerates with time according to
$r [p(t)]^{\delta}$.
The generic consequence of a power law acceleration in the growth rate is the
appearance of singularities in finite time:
\be \label{pow}
p(t) \propto (t_c - t)^z, ~~{\rm with}~z=-{1 \over \delta}~~ \mbox{ and $t$
close to $t_c$}.
\ee
Equation (\ref{aafajdak}) is said to have a ``spontaneous'' or ``movable''
singularity at the critical time $t_c$ \cite{benderorszag}, the critical time
$t_c$ being determined by the constant of integration, {\it i.e.}, the initial
condition $p(t=0)$. One can get an intuitive  understanding of such
singularities by looking at the function $p(t) = \exp \left( t p(t) \right)$
which corresponds to
replacing $\tau K$ by $p$ in Malthus' exponential solution
$p(t) = p(0) \exp [\tau K t]$. $p$ is then the solution of
$dp/dt = p^2/(1 - t p)$ \cite{benderorszag}
leading to an ever increasing growth  with the explicit solution
\be
p(t) = e \left( 1 - C \sqrt{t_c - t}\right)~,
\ee
where $t_c=1/e=0.368$, $C$ is a numerical factor and the exponent $z=1/2$. In
this case, the finite time spontaneous singularity does not lead to a
divergence of the population at the critical time $t_c$; only the growth
rate diverges at $(t_c-t)^{-1/2}$. Spontaneous singularities in ODE's and
PDE's are quite common and have been found in many well-established models
of natural systems either at special points in space such as in the Euler
equations of inviscid fluids \cite{Pumiersiggia} or in the equations of
General Relativity coupled to a mass field leading to the formation of
black holes \cite{Choptuik}, in models of micro-organisms aggregating to form
fruiting bodies \cite{Rascle}, or to the more prosaic rotating coin (Euler's
disk) \cite{Moffatt}, see \cite{reviewsor} for a review. Some of the
most prominent, as well as more controversial, examples due to their impact on
human society are models of rupture and material failure \cite{failure,faicri},
earthquakes \cite{earthquake} and stock market crashes \cite{crash,nasdaq}.

\subsection{Data sets and methodology}

Here, we examine several data sets expressing the
development of mankind on Earth
in term of size and economic impact,
to test the hypothesis that our history might be compatible with a
future finite-time singularity. These data sets are as follows.
\begin{itemize}
\item The human population data from 0 to 1998 was retrieved from the
web-site of The United Nations Population Division
Department of Economic and Social Affairs (http://www.popin.org/pop1998/).

\item The GDP of the world from 0 to 1998, estimated by J. Bradford DeLong
at the Department of Economics, U.C. Berkeley \cite{Bradford}, was
given to us by R. Hanson \cite{Hanson}.

\item The financial data series include the Dow Jones index from 1790
\cite{UShistory} to 2000, the Standard \& Poor (S\&P) index from 1871 to 2000,
as well as a number of regional and global indices since 1920. The Dow Jones
index was constructed by The Foundation for the Study of Cycles \cite{cycles}.
It is the Dow Jones index back to 1896, which has been extrapolated back to
1790 and further. The other indices are from Global Financial Data
\cite{global}.
These indices are constructed as follows. For the S\&P, the data from 1871
to 1918
are from the Cowles commission, which back-calculated the data using the
{\it Commercial and  Financial Chronicle}. From 1918, the data is the Standard
and Poor's Composite index (S\&P) of stocks. The other indices uses Global
Financial Data's indices from 1919 through 1969 and Morgan Stanley Capital
International's indices from 1970 through 2000. The EAFE Index includes
Europe,
Australia and the Far East. The Latin America Index includes Argentina,
Brazil,
Chile, Colombia, Mexico, Peru and Venezuela.
\end{itemize}

Demographers usually construct population projections in 
a disaggregated manner, filtering the data by age, stage of development, region, etc.
Disaggregating and controlling for such variables are thought to be crucial for
demographic development and for any reliable population prediction.
Here, we propose a different strategy based on aggregated data, which is
justified by the following concept: in order to get a meaningful prediction
at an aggregate level, it is often more relevant to study aggregate
variables than ``local'' variables that can miss the whole picture in 
favor of special idiosyncrasies.
 To take an example from material sciences, the prediction of the
failure of heterogeneous materials subjected to stress can be performed
according to two methodologies. Material scientists often analyse in 
exquisite details the wave forms of the acoustic emissions or other
signatures of damage resulting from
micro-cracking within the material. However, this is of very little help
to predict the overall failure which is often a cooperative global
phenomenon \cite{Herroux} resulting from 
the interactions and interplay between the 
many different micro-cracks nucleating, growing and fusing within the 
materials. In this example, it has been shown indeed that aggregating
all the acoustic emissions in a single aggregated variable is much better
for prediction purpose \cite{faicri}.

\subsection{Content of the paper}

In the next section, we first show that the exponential model is utterly
inadequate in
describing
the population growth as well as the growth in the World GDP and the global and
regional financial indices. We then present the alternative model
consisting of a power law
growth ending at a critical time $t_c$. We first give a non-parametric approach
complemented by a fitting procedure. Section 3 proposes a first generalization
of power laws with complex exponents, leading to so-called log-periodic
oscillations
decorating the overall power law acceleration. The fitting procedure is
described
as well as a non-parametric test of the existence of the log-periodic
patterns for
the world population. Section 4 presents a second-order generalization of
the power law
model, which allows for a frequency modulation in the log-periodic structure.
This extended formula is used to fit the extended Dow Jones Industrial average.
Section 5 summarizes what has been achieved and compares our results with
previous work.
In particular, we give the explicit solutions of multivariate dynamical
equations for
several coupled variables, such as population, technology and capital, to
show that
the same finite-time singularity can emerge from the interplay of these
factors while
each of them individually is not enough to create the singularity. Section
6 concludes by
discussing a set of scenarios for mankind close to and beyond the critical
time.

\section{Singular Growth Rate}

\subsection{Tests of exponential growth}

\subsubsection{Human population and world GDP}

\begin{figure}
\begin{center}
\parbox[l]{8.5cm}{
\epsfig{file=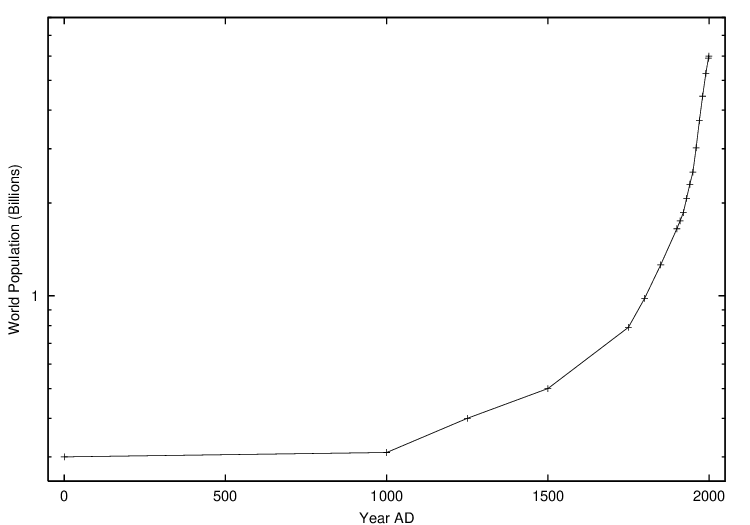,height=8cm,width=8.5cm}
\caption{\label{semilogpop}Semi-logarithmic plot of World population
from year 0 until Oct. 1999.
In this representation, a linear increase would qualify an exponential
growth. Note in contrast
the super-exponential behavior.}}
\hspace{5mm}
\parbox[r]{8.5cm}{
\epsfig{file=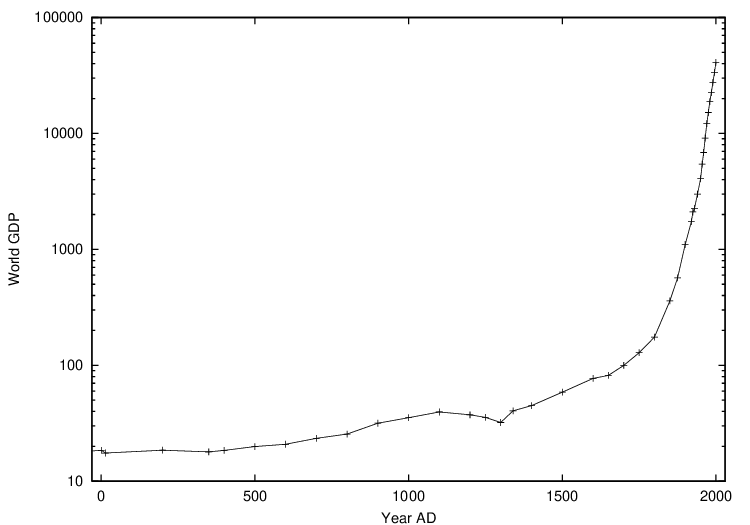,height=8cm,width=8.5cm}
\caption{\label{semilogwgdp}Semi-logarithmic plot of World GDP from
year 1 until 2000.}}

\vspace{1.5cm}

\parbox[l]{8.5cm}{
\epsfig{file=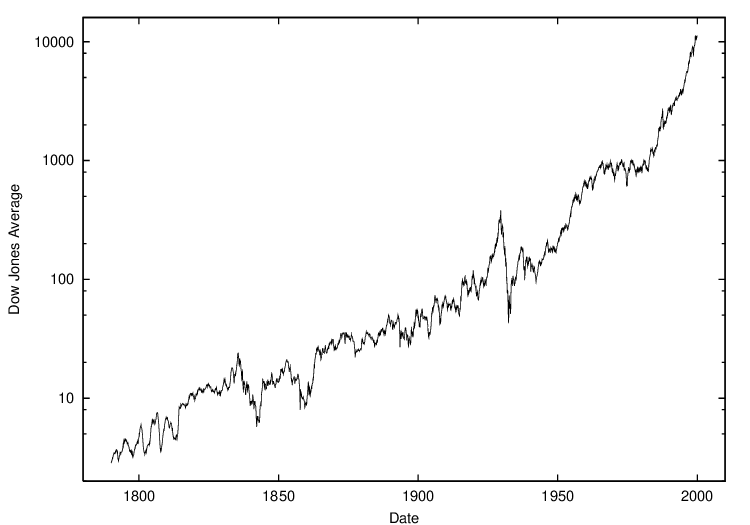,height=8cm,width=8.5cm}
\caption{\label{semilogdj}Semi-logarithmic plot of the Dow Jones from 1790
until 2000.}}
\hspace{5mm}
\parbox[r]{8.5cm}{
\epsfig{file=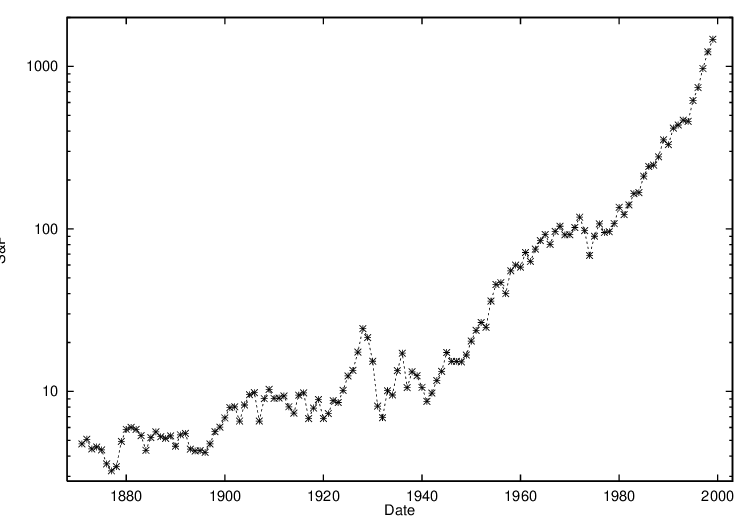,height=8cm,width=8.5cm}
\caption{\label{semilogsp}Semi-logarithmic plot of the S\&P from 1871 until
2000. }}
\end{center}
\end{figure}

A faster than exponential growth is clearly observed in the human population
data from year 0 up to 1970, at which the estimated annual rate of
increase of
the global population reached its (preliminary?) all-time peak of $2.1\%$.
Figure \ref{semilogpop} shows the logarithm of the estimated world population
as a function of (linear) time, such that an exponential growth rate would be
qualified by a linear increase. In contrast, one clearly observes a strong
upward
curvature characterising a ``super-exponential'' behaviour.
A faster than exponential growth is also clearly observed
in the estimated GDP (Gross Domestic Product) of the World,
shown in figure \ref{semilogwgdp} for the year 0 up to 2000.

\subsubsection{Financial indices}

\begin{figure}
\begin{center}
\parbox[l]{8.5cm}{
\epsfig{file=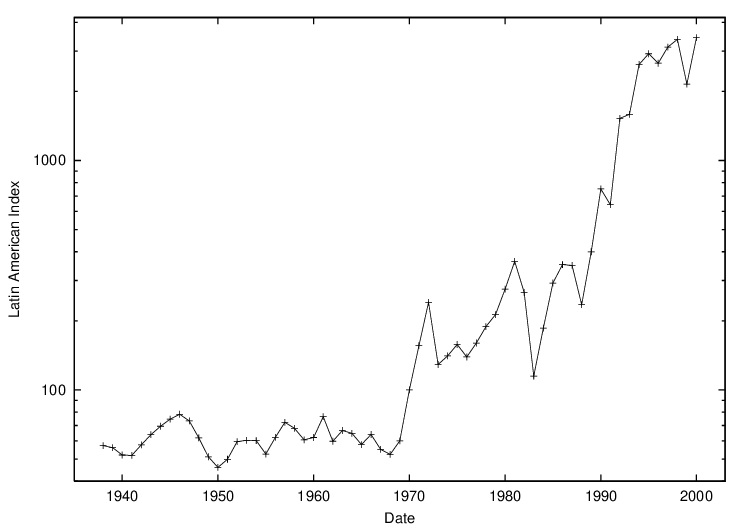,height=8cm,width=8.5cm}
\caption{\label{semiloglat}Semi-logarithmic plot of the Latin American
index from 1938 until 2000.}}
\hspace{5mm}
\parbox[r]{8.5cm}{
\epsfig{file=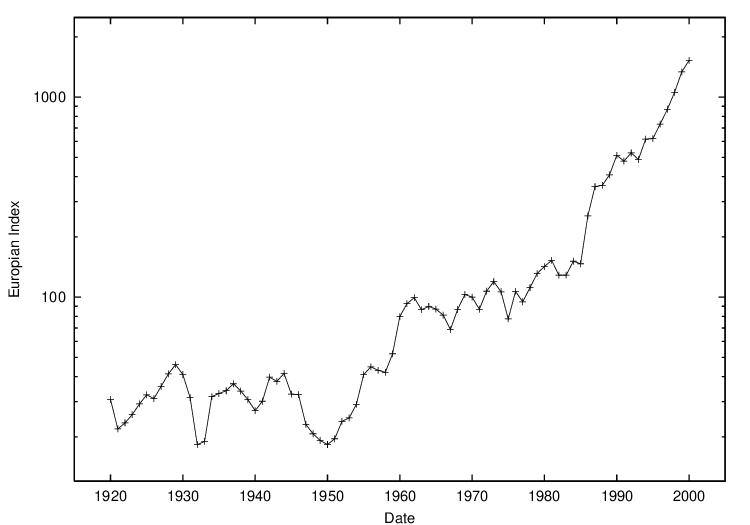,height=8cm,width=8.5cm}
\caption{\label{semilogeur}Semi-logarithmic plot of the European index from
1920 until 2000.}}

\vspace{1.5cm}

\parbox[l]{8.5cm}{
\epsfig{file=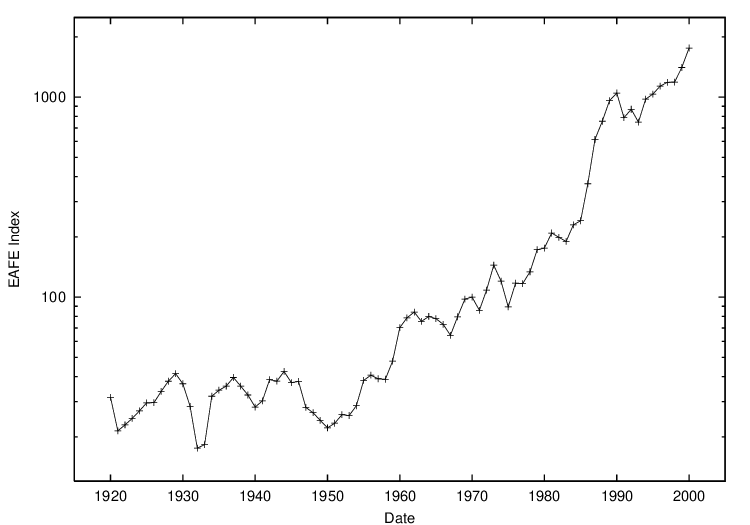,height=8cm,width=8.5cm}
\caption{\label{semilogeafe}Semi-logarithmic plot of the EAFE index from
1920 until 2000.}}
\hspace{5mm}
\parbox[r]{8.5cm}{
\epsfig{file=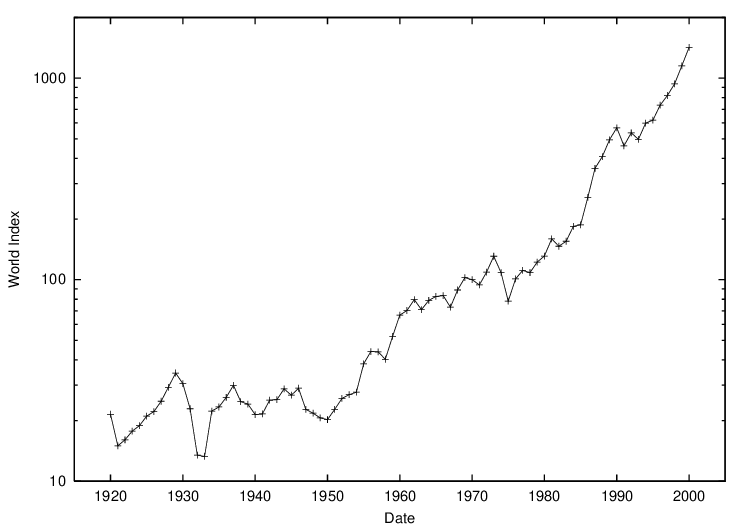,height=8cm,width=8.5cm}
\caption{\label{semilogwindex}Semi-logarithmic plot of the World index from
1920 until 2000.}}
\end{center}
\end{figure}

Over a shorter time period, a faster than exponential growth is also
observed in figures \ref{semilogdj} to
\ref{semilogwindex} for a number of
economic indicators such as the Dow Jones Average since the establishment
of the
U.S.A. in 1790 \cite{UShistory}, the S\&P since 1871, as well for a number
of regional and global indices since 1920, including the Latin American
index, the
European index, the EAFE index and the World index. In all these figures,
the logarithm of the index is plotted
as a function of (linear) time, such that an exponential growth rate would be
qualified by a linear increase. In all cases, one clearly observes in contrast
a significant upward
curvature characterising a ``super-exponential'' behaviour.

\subsection{A first test of power law growth}

\subsubsection{Procedure}

As shown in the derivation of equation (\ref{aafajdak}), it is enough that
the growth rate
increases with any arbitrarily small {\it positive} power of $p(t)$ for
a finite-time singularity to develop with the characteristic power law
dependence
(\ref{pow}). Can such a behaviour explain the super-exponential behaviour
documented in the figures \ref{semilogpop}-\ref{semilogwindex}?

The small number of data points in these time series
and the presence of large fluctuations prevent the use of a direct fitting
procedure with
(\ref{pow}). Indeed, such a fit, which typically attempts to minimise the
root-mean-square (r.m.s.)
difference
between the theoretical formula and the data, is highly degenerate: many
solutions are found which differ by variations of at most a few percent of
the root-mean-square (r.m.s.) of the errors. Such
differences in r.m.s. are not significant, especially considering the
strongly non-Gaussian nature of the fluctuations in these data sets. Maximum
likelihood methods
are similarly limited. To address this problem of degeneracy, we turn to a
non-parametric approach,
consisting in fixing $t_c$ and plotting the logarithm of the data as a
function of
$\log (t_c -t)$. In such a plot, a linear behaviour qualifies the power law
(\ref{pow}),
and the slope gives the exponent $z$ which then can be
determined visually or, better, by a fit but now with $t_c$ fixed. This
procedure
is not plagued by the previously discussed degeneracy and provides reliable
and unique results.

\subsubsection{World population}

\begin{figure}
\begin{center}
\parbox[l]{8.5cm}{
\epsfig{file=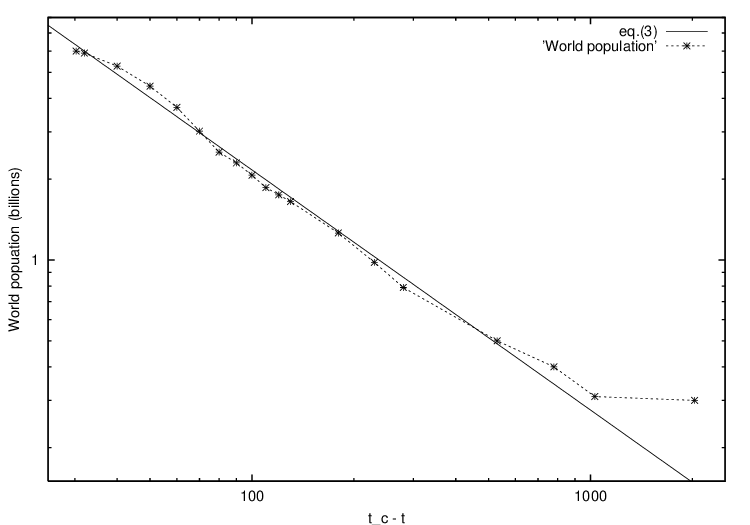,height=8cm,width=8.5cm}
\caption{\label{pop2030} World population as a function of $t_c - t$
with $t_c=2030$.
The straight line is the
fit with a power law $p(t) = a(t_c - t)^z$ with a fixed $t_c=2030$, see
Table \protect\ref{ztable}.}}
\hspace{5mm}
\parbox[r]{8.5cm}{
\epsfig{file=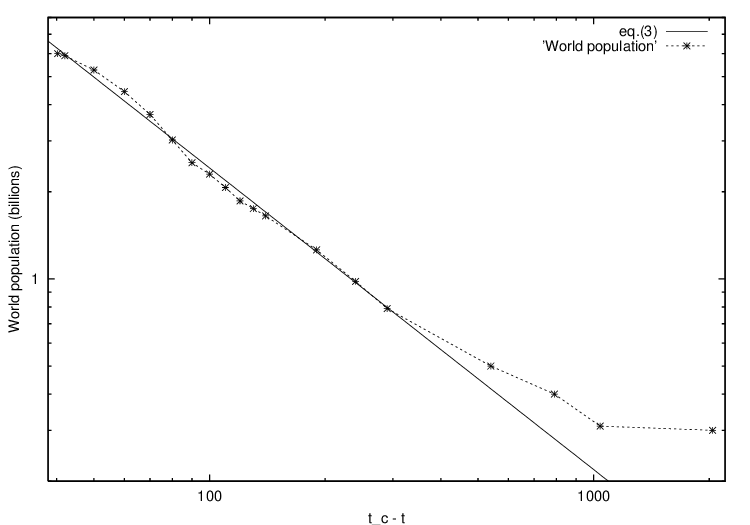,height=8cm,width=8.5cm}
\caption{\label{pop2040} World population as a function of $t_c - t$
with $t_c=2040$.
The straight line is the
fit with a power law $p(t) = a(t_c - t)^z$ with $t_c=2040$ fixed, see Table
\protect\ref{ztable}.}}

\vspace{1.5cm}

\parbox[l]{8.5cm}{
\epsfig{file=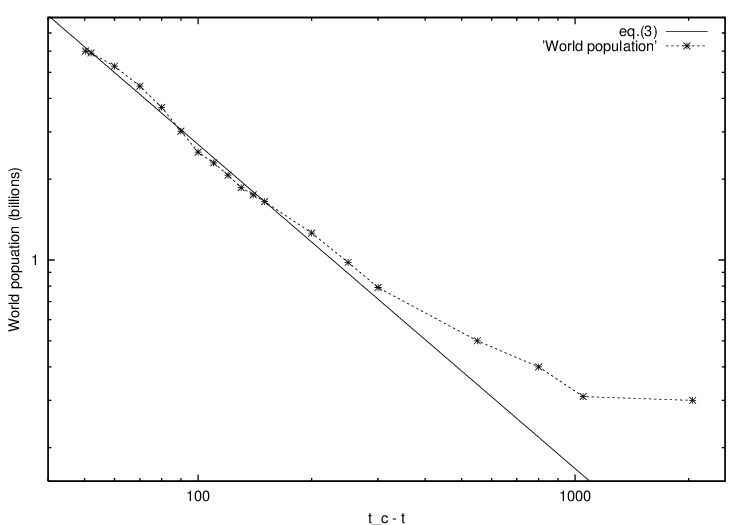,height=8cm,width=8.5cm}
\caption{\label{pop2050} World population as a function of $t_c - t$
with $t_c=2050$.
The straight line is the
fit with a power law $p(t) = a(t_c - t)^z$ with $t_c=2050$ fixed, see Table
\protect\ref{ztable}.}}
\hspace{5mm}
\parbox[r]{8.5cm}{
\epsfig{file=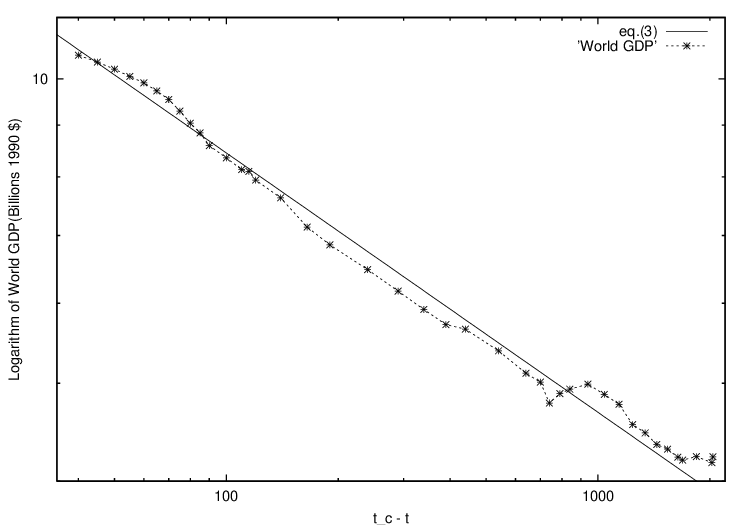,height=8cm,width=8.5cm}
\caption{\label{wgdp2040} World GDP as a function of $t_c - t$ with
$t_c=2040$.
The straight line is the
fit with a power law $p(t) = a(t_c - t)^z$ with $t_c=2040$ fixed, see Table
\protect\ref{ztable}.}}
\end{center}
\end{figure}

In figures \ref{pop2030}-\ref{pop2050},
the world population in logarithm scale is shown as a function of $t_c - t$
also in
logarithmic scale for three choices 2030, 2040 and 2050, respectively for
$t_c$.
Even though the fits
with equation (\ref{pow}) for three cases varies in quality, they all
capture the acceleration
in the second half of the data on a logarithmic scale. The curvature seen
in the data far from $t_c$ can be modeled by including a constant
term in equation (\protect\ref{pow}) embodying for instance the effect of
an initial
condition, as we discuss below. Changing $t_c$ from 2030 to 2050
has two competing effects observed in figures \ref{pop2030}-\ref{pop2050}:
a larger value
of $t_c$ provides a better fit in the latter time period while
deteriorating somewhat the
fit to the data in the early time periods.

\subsubsection{World GDP}

\begin{figure}
\begin{center}
\parbox[l]{8.5cm}{
\epsfig{file=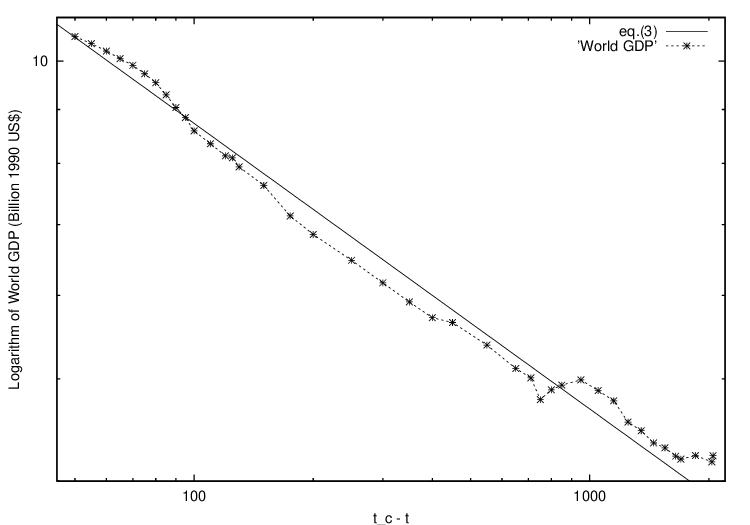,height=8cm,width=8.5cm}
\caption{\label{wgdp2050} World GDP as a function of $t_c - t$ with
$t_c=2050$.
The straight line is the
fit with a power law $p(t) = a(t_c - t)^z$ with $t_c=2050$ fixed, see Table
\protect\ref{ztable}.}}
\hspace{5mm}
\parbox[r]{8.5cm}{
\epsfig{file=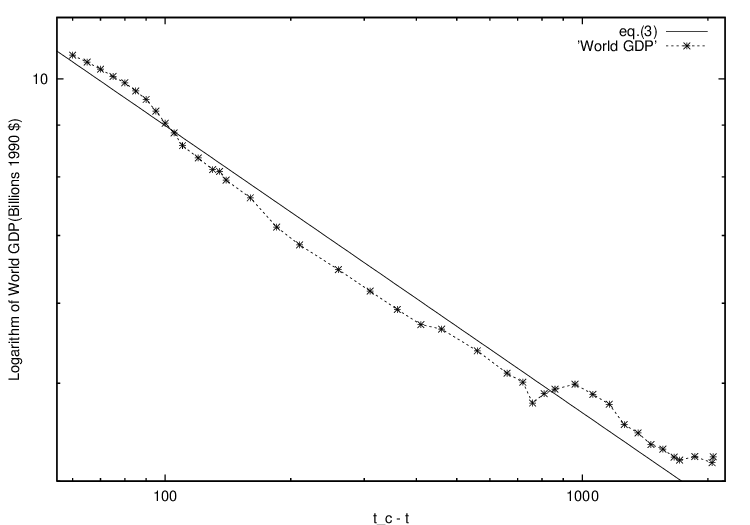,height=8cm,width=8.5cm}
\caption{\label{wgdp2060}World GDP as a function of $t_c - t$ with
$t_c=2060$.
The straight line is the
fit with a power law $p(t) = a(t_c - t)^z$ with $t_c=2060$ fixed, see Table
\protect\ref{ztable}.}}

\vspace{1.5cm}

\parbox[l]{8.5cm}{
\epsfig{file=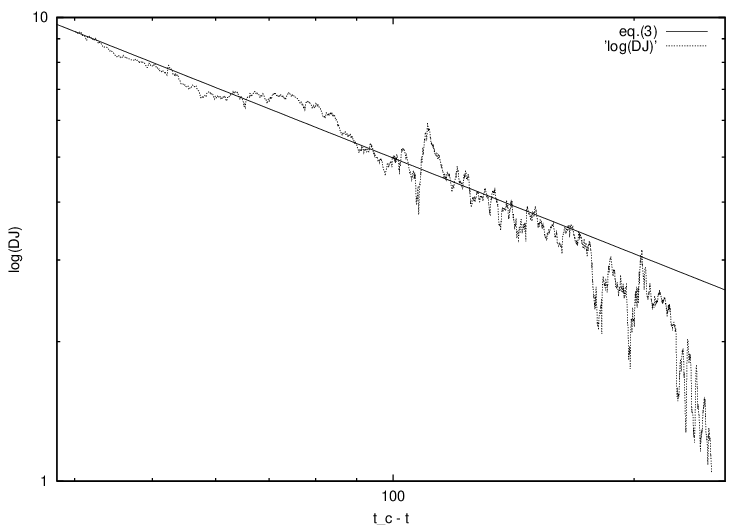,height=8cm,width=8.5cm}
\caption{\label{dj2040}The Dow Jones as a function of $t_c - t$ with
$t_c=2040$.
The straight line is the
fit with a power law $p(t) = a(t_c - t)^z$ with $t_c=2040$ fixed, see Table
\protect\ref{ztable}.}}
\hspace{5mm}
\parbox[r]{8.5cm}{
\epsfig{file=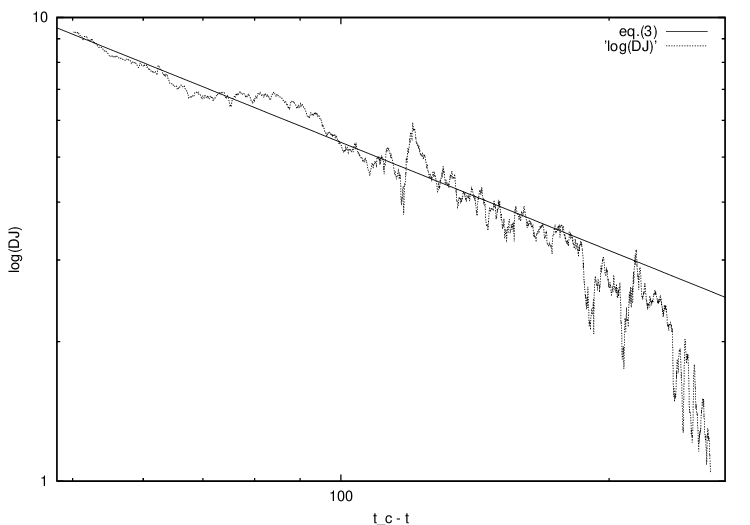,height=8cm,width=8.5cm}
\caption{\label{dj2050}The Dow Jones as a function of $t_c - t$ with
$t_c=2050$.
The straight line is the
fit with a power law $p(t) = a(t_c - t)^z$ with $t_c=2050$ fixed, see Table
\protect\ref{ztable}.}}
\end{center}
\end{figure}

As discussed in the introduction, the human population is strongly coupled
with its outputs and with the Earth's carrying capacity, and can partly be
measured by its
economic production. Hence, we should expect a close relationship between
the size of the human population and its GDP.
Figures \ref{wgdp2040}-\ref{wgdp2060} show the logarithm
of the estimated World GDP as a function of
 $t_c - t$, both in log-log coordinates, where $t_c$ has been chosen to
2040, 2050
and 2060, respectively. The equation (\ref{pow}) is again parameterising
the data
quite satisfactorily.

We stress that we use the logarithm of the World
GDP as well as the logarithm of the national, regional or global indices
presented below
as the ``bare'' data on which we test the power law hypothesis. This means
that we plot
the logarithm of the GDP or of the indices in logarithmic scale,
which effectively amounts to taking
the logarithm of the logarithm of the GDP as a function of $t_c-t$,
itself also
 in logarithmic scale in order to test for the power law (\ref{pow}).
This is done in an attempt to minimise the effect of inflation and other
systematic drifts,
and in accordance with standard economic practice that only
relative changes should be considered. Removing an average inflation of 4\%
does not change
the results qualitatively but the corresponding results are not
quantitatively reliable
as the inflation has varied
significantly over US history with quantitative impacts that are difficult
to estimate.

\subsubsection{Financial indices}

\begin{figure}
\begin{center}
\parbox[l]{8.5cm}{
\epsfig{file=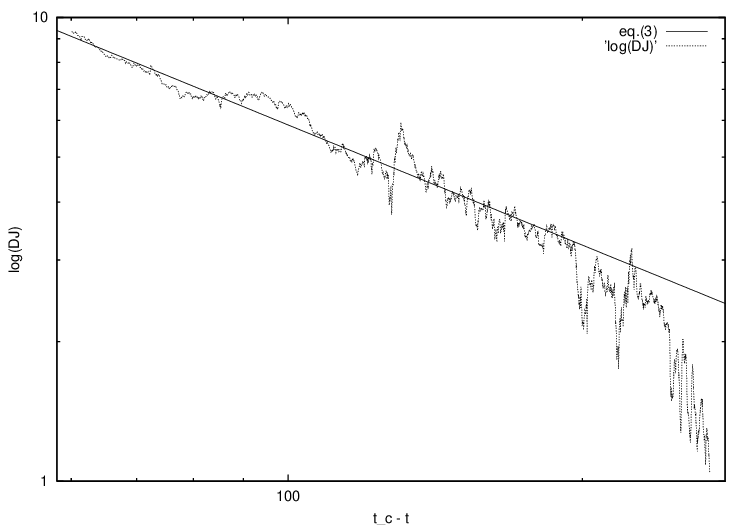,height=8cm,width=8.5cm}
\caption{\label{dj2060}The Dow Jones as a function of $t_c - t$ with
$t_c=2060$.
The straight line is the
fit with a power law $p(t) = a(t_c - t)^z$ with $t_c=2060$ fixed, Tsee able
\protect\ref{ztable}.}}
\hspace{5mm}
\parbox[r]{8.5cm}{
\epsfig{file=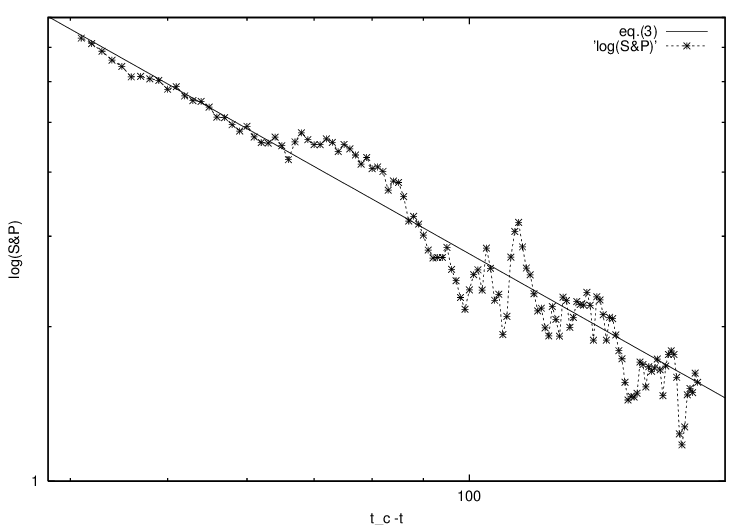,height=8cm,width=8.5cm}
\caption{\label{sp2040}The S\&P as a function of $t_c - t$ with $t_c=2040$.
The straight line is the
fit with a power law $p(t) = a(t_c - t)^z$ with $t_c=2040$ fixed, see Table
\protect\ref{ztable}.}}

\vspace{1.5cm}

\parbox[l]{8.5cm}{
\epsfig{file=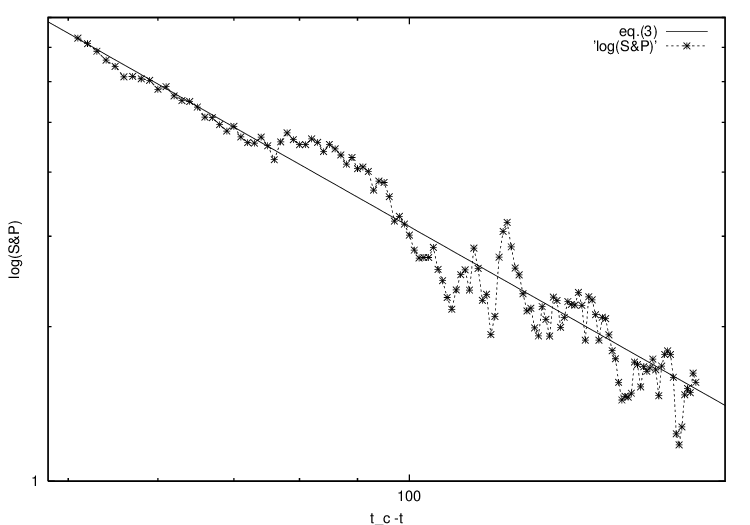,height=8cm,width=8.5cm}
\caption{\label{sp2050}The S\&P as a function of $t_c - t$ with $t_c=2050$.
The straight line is the
fit with a power law $p(t) = a(t_c - t)^z$ with $t_c=2050$ fixed, see Table
\protect\ref{ztable}.}}
\hspace{5mm}
\parbox[r]{8.5cm}{
\epsfig{file=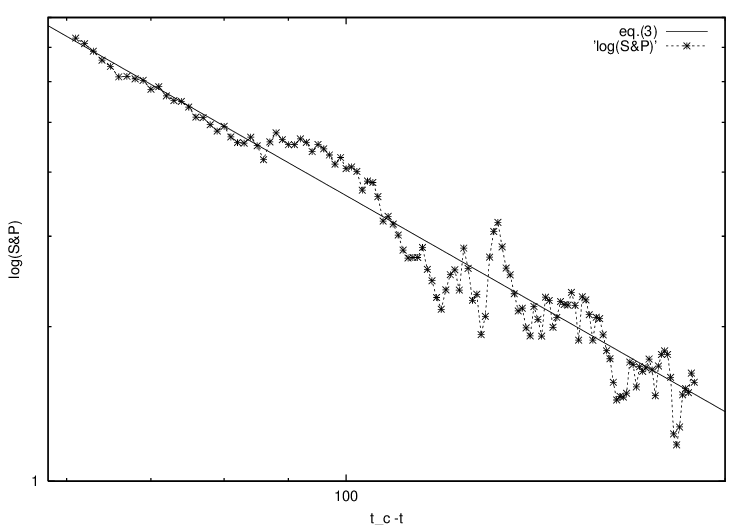,height=8cm,width=8.5cm}
\caption{\label{sp2060}The S\&P as a function of $t_c - t$ with $t_c=2060$.
The straight line is the
fit with a power law $p(t) = a(t_c - t)^z$ with $t_c=2060$ fixed, see Table
\protect\ref{ztable}.}}
\end{center}
\end{figure}

\begin{figure}
\begin{center}
\parbox[l]{8.5cm}{
\epsfig{file=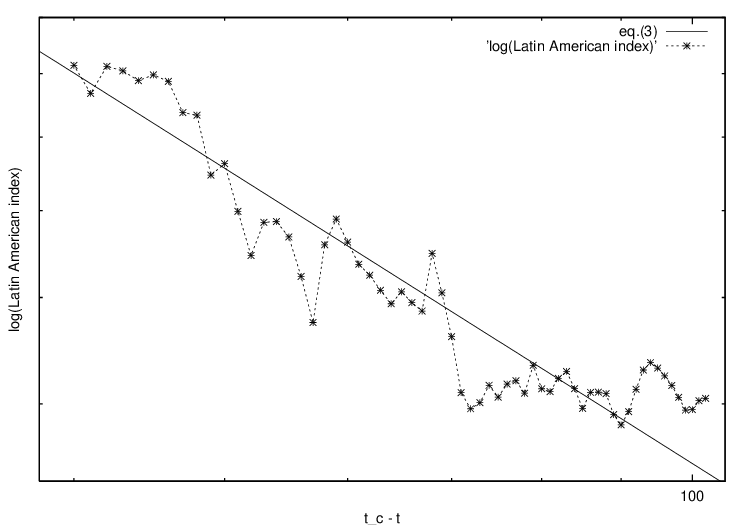,height=8cm,width=8.5cm}
\caption{\label{lat2040}The Latin America index as a function of $t_c - t$
with
$t_c=2040$. The straight line is the
fit with a power law $p(t) = a(t_c - t)^z$ with $t_c=2040$ fixed, see Table
\protect\ref{ztable}.}}
\hspace{5mm}
\parbox[r]{8.5cm}{
\epsfig{file=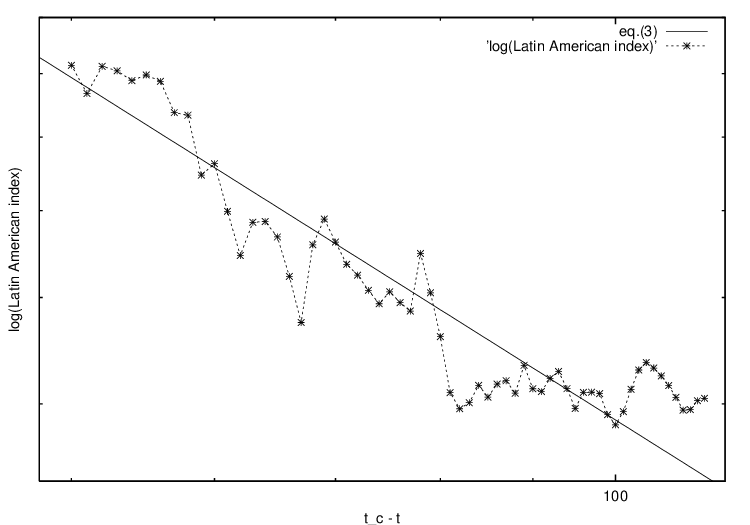,height=8cm,width=8.5cm}
\caption{\label{lat2050}The Latin America index as a function of $t_c - t$ with
$t_c=2050$. The straight line is the
fit with a power law $p(t) = a(t_c - t)^z$ with $t_c=2050$ fixed, see
Table \protect\ref{ztable}.}}

\vspace{1.5cm}

\parbox[l]{8.5cm}{
\epsfig{file=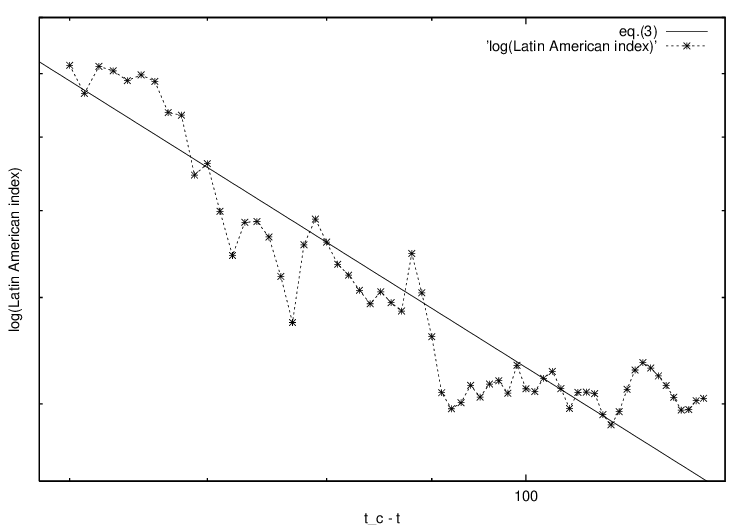,height=8cm,width=8.5cm}
\caption{\label{lat2060}The Latin America index as a function of $t_c - t$
with $t_c=2060$.
The straight line is the
fit with a power law $p(t) = a(t_c - t)^z$ with $t_c=2060$ fixed, see Table
\protect\ref{ztable}.}}
\hspace{5mm}
\parbox[r]{8.5cm}{
\epsfig{file=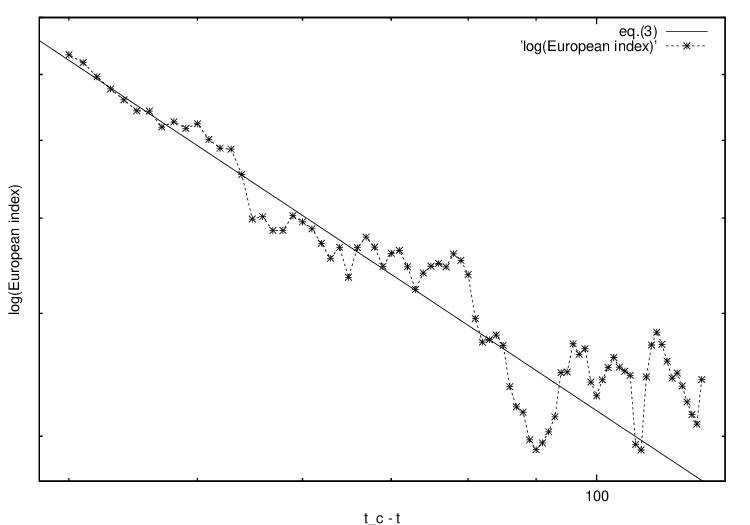,height=8cm,width=8.5cm}
\caption{\label{eur2040}The European index as a function of $t_c - t$ with
$t_c=2040$.
The straight line is the
fit with a power law $p(t) = a(t_c - t)^z$ with $t_c=2040$ fixed, see Table
\protect\ref{ztable}.}}
\end{center}
\end{figure}

\begin{figure}
\begin{center}
\parbox[l]{8.5cm}{
\epsfig{file=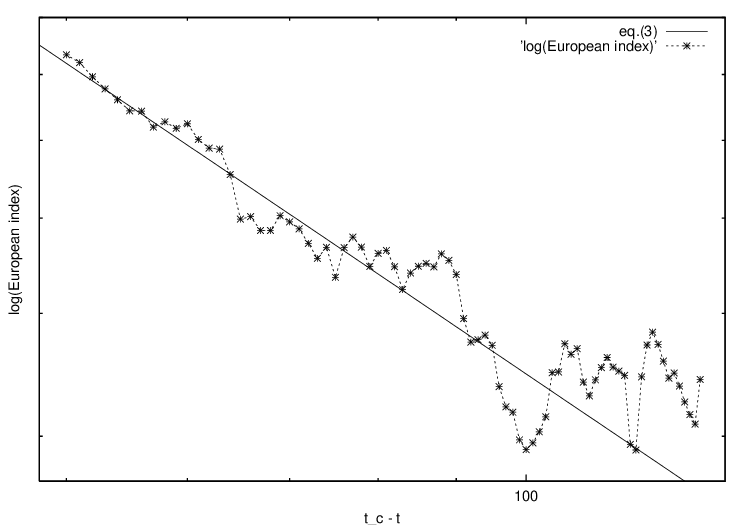,height=8cm,width=8.5cm}
\caption{\label{eur2050}The European index as a function of $t_c - t$ with
$t_c=2050$.
The straight line is the
fit with a power law $p(t) = a(t_c - t)^z$ with $t_c=2050$ fixed, see Table
\protect\ref{ztable}.}}
\hspace{5mm}
\parbox[r]{8.5cm}{
\epsfig{file=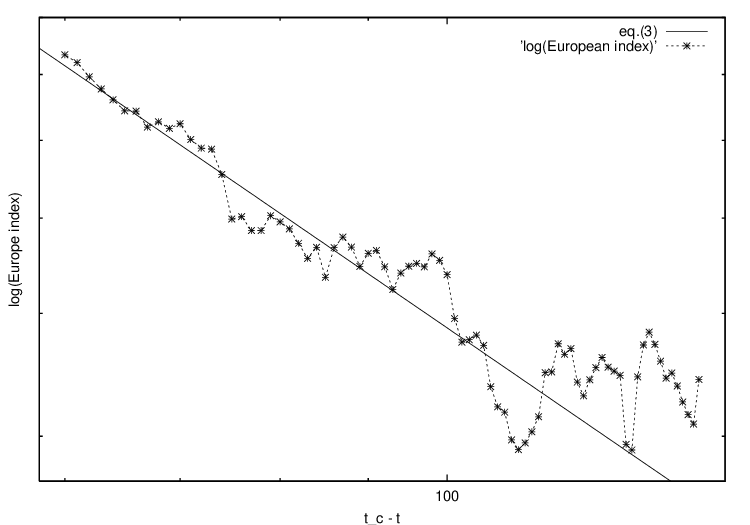,height=8cm,width=8.5cm}
\caption{\label{eur2060}The European index as a function of $t_c - t$ with
$t_c=2060$.
The straight line is the
fit with a power law $p(t) = a(t_c - t)^z$ with $t_c=2060$ fixed, see Table
\protect\ref{ztable}.}}

\vspace{1.5cm}

\parbox[l]{8.5cm}{
\epsfig{file=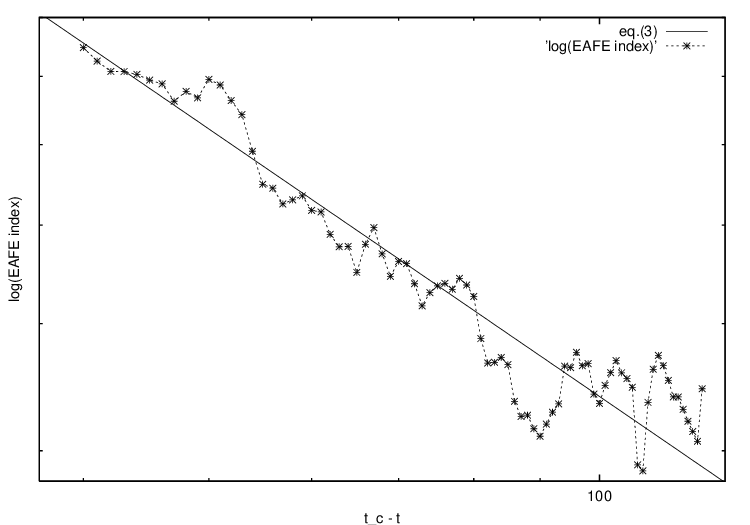,height=8cm,width=8.5cm}
\caption{\label{eafe2040}The EAFE index as a function of $t_c - t$ with
$t_c=2040$.
The straight line is the
fit with a power law $p(t) = a(t_c - t)^z$ with $t_c=2040$ fixed, see Table
\protect\ref{ztable}.}}
\hspace{5mm}
\parbox[r]{8.5cm}{
\epsfig{file=fitlogeur2050.eps,height=8cm,width=8.5cm}
\caption{\label{eafe2050}The EAFE index as a function of $t_c - t$ with
$t_c=2050$.
The straight line is the
fit with a power law $p(t) = a(t_c - t)^z$ with $t_c=2050$ fixed, see Table
\protect\ref{ztable}.}}
\end{center}
\end{figure}

Further support for a singular power law behaviour of the economy
can be found by analysing in a similar way the national, regional or global
indices shown in figures
\ref{semilogdj}-\ref{semilogwindex}. The results are shown in figures
\ref{dj2040}-\ref{windex2060}.
Equation (\ref{pow}) is again
perfectly compatible with the data and much better than any exponential
model.
As shown in Table \ref{ztable}, the fits of all six indices are found to be
consistent with similar
values for the exponent $z \approx -1$, the absolute value of
the exponent increasing with $t_c$.

The results presented in this section on the world population, on the world
GDP and on
six financial indices suggest that the power law (\ref{pow}) is an adequate
model. It is also
parsimonious since the same simple mathematical expression, approximately
the same critical time
$t_c$ and same exponent are found consistently for all time series.
These results confirm and extend the analysis presented forty years earlier
for the
world population only
\cite{Doomsdaypaper}, which concluded at a $t_c=2026$. The results shown in
the figures
\ref{pop2030}-\ref{pop2050}, with the sensitivity analysis provided by
varying $t_c$ from 2030 to 2050, illustrate the large uncertainty in its
determination.
It is thus worthwhile to attempt quantifying further the observed power law
growth and
test how well $t_c$ is constrained.

\subsection{Quantitative fits to a power law}

In the derivation of (\ref{pow}), a key assumption was to neglect the
limiting negative
term in (\ref{ajdak}), which is warranted sufficiently close to $t_c$. Far
from $t_c$, this
analysis and more general considerations lead us to expect the existence of
corrections to the pure
power law (\ref{pow}). Furthermore, it may be necessary to include
higher order terms as well as generalise the exponent as we will see in the
next
section.

\begin{figure}
\begin{center}
\epsfig{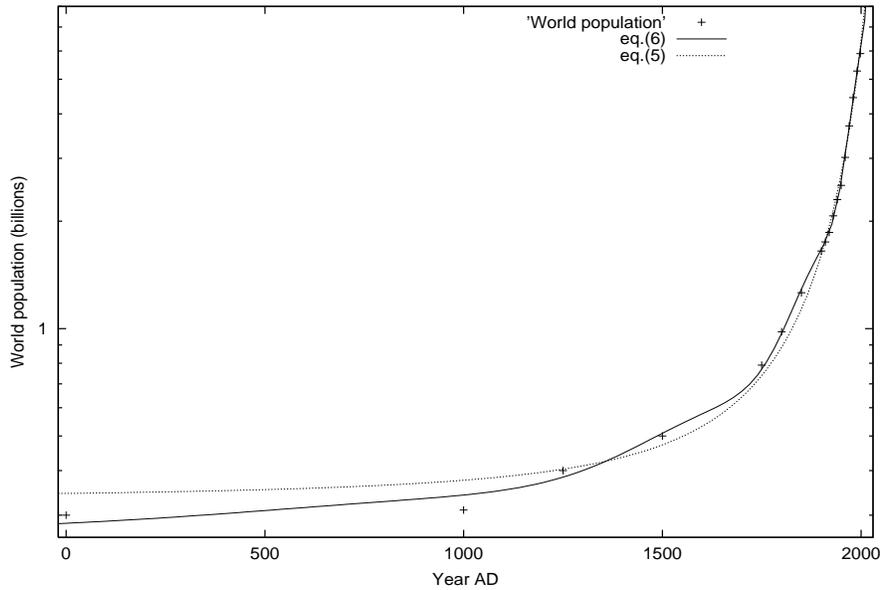}
\caption{\protect\label{powlpfits} The dotted line is the best fit with
equation $(\protect\ref{eq:solution})$ to data set 5, see text. The fit
gives $\mbox{r.m.s.}=0.111$, $A\approx 0$, $B \approx 22120$, $t_c \approx
2078$
and $z \approx -1.9$. The full line is the best fit with equation
$(\protect\ref{lppow})$ and gives $\mbox{r.m.s.}=0.030$, $A\approx 0$,
$B\approx 1624$, $C\approx -127$, $z\approx -1.4$, $t_c \approx 2056$, $\omega
\approx 6.3$ and $\phi \approx 5.1$.}
\end{center}
\end{figure}

The simplest extension of equation (\ref{pow}) is
\be
\label{eq:solution}
p(t) = A  + B (t_c-t)^{z}~.
\ee
In order to make a first quantitative estimate of the acceleration in the
growth rate, determined
by the exponent $z$ and the position $t_c$ of the singularity,
we now let $t_c$ be a free parameter. In figure \ref{powlpfits},
the equation (\ref{eq:solution}) is fitted to the world
population from 0 to 1998.  The parameter values of the fit are
$A\approx 0$, $B \approx 22120$, $t_c \approx 2078$ and $z \approx -1.9$.
The negative value of the exponent is compatible with $A\approx 0$. The
negative
exponent $z \approx -1.9$ obtained in the fit means that equation
(\ref{eq:solution})
has a singularity at $t=t_c$ corresponding to an infinite population. This is
clearly impossible on a finite Earth. The point to be extracted from this
analysis
is that the world population has until very recently grown
at an accelerating growth rate in good agreement with a singular behaviour.
Singularities are always mathematical idealisations of natural phenomena:
they are
not present in reality but foreshadow an important transition or change of
regime.
In the present context, they must be interpreted as a kind of ``critical
point'' signaling a fundamental and abrupt change of regime similar to what
occurs
in phase transitions \cite{critical}.

\begin{figure}
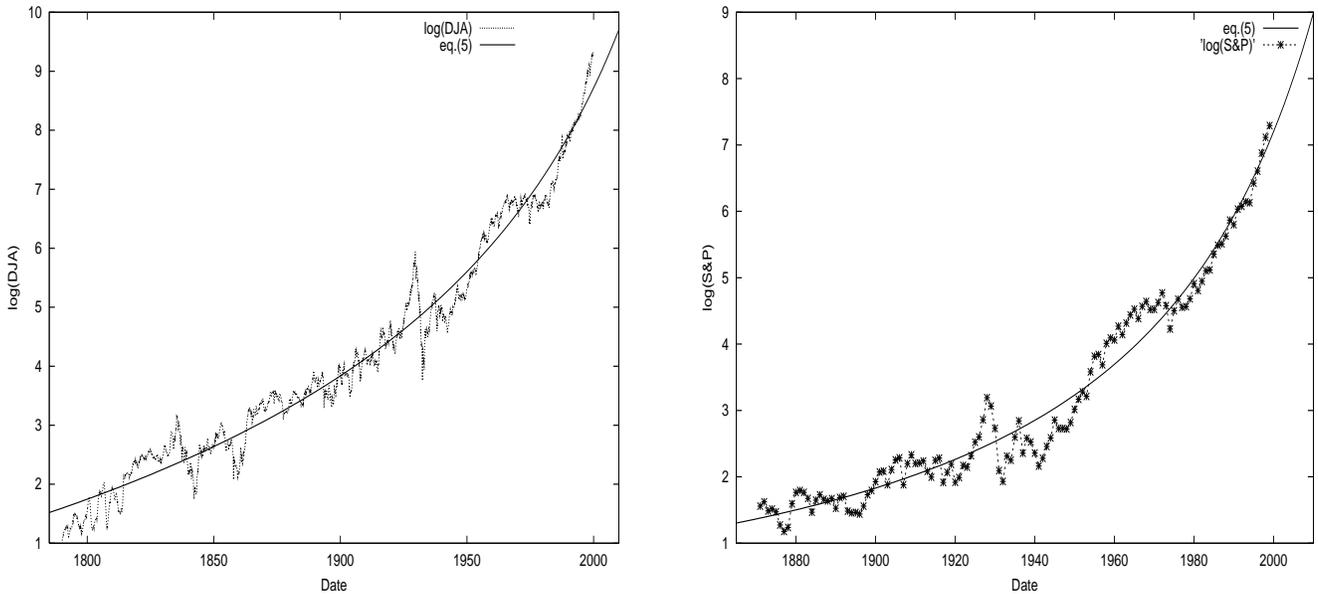

\begin{center}
\parbox[l]{8.5cm}{
\epsfig{file=11-2nspow252001logdj1790.eps,height=8cm,width=8.5cm}}
\hspace{5mm}
\parbox[r]{8.5cm}{
\epsfig{file=fitpowlogsp1871.eps,height=8cm,width=8.5cm}}
\caption{\label{powlogdjsp}Left: The Dow Jones Average fitted with equation
\protect\ref{eq:solution}. The values of the fit are $A\approx-14$,
$B\approx 71$,
$z\approx-0.27$ and $t_c\approx2068$. Right: The S\&P fitted with equation
\protect\ref{eq:solution}. The values of the fit are $A\approx 0$,
$B\approx 1693$,
$z\approx -1.3$ and $t_c\approx 2067$.}
\end{center}
\end{figure}

As already discussed in relation with equation (\ref{ajdak}) and in the
previous section, the world population growth cannot
be separated from that of its evolving carrying
capacity. As a first attempt to quantify this variable in an independent way,
we analyse quantitatively the two largest data sets among all the
financial indices and GDP: due to the large fluctuations of the financial
indices compared to
the number of points, only the S\&P and the Dow Jones
Average gave reliable results when $t_c$ is a free parameter.
Figure \ref{powlogdjsp} shows
the corresponding fits with equation (\ref{eq:solution}). The parameter values
of the best fits are $A\approx-14$, $B\approx 71$, $z\approx-0.27$ and
$t_c\approx 2068$ for the Dow Jones Average and $A\approx 0$, $B\approx 1693$,
$z\approx-1.3$ and $t_c\approx 2067$ for the S\&P.
The fit with equation (\ref{eq:solution}) exemplifies the acceleration of the
growth rate, which is our main message. However, the location of the critical
point is still not very reliable when based on simple power fits
of very noisy data \cite{DombGreen}. This motivates us to extend this analyses
in the following sections.

\section{Beyond a simple power law}

The results shown in the figures
\ref{pop2030}-\ref{windex2060}, with the sensitivity analysis provided by
varying $t_c$ from 2030 or 2040 to 2050 or 2060, illustrate the large
uncertainty in the determination of the critical time. The direct fit with
(\ref{eq:solution}) still gives a very large uncertainty. As can be seen
from the figures, an important reason lies in the existence of large
fluctuations around the average power law behaviour. In the next section,
we will see that this variability might be genuine and not simply noise.
Furthermore, adding an extra degree of freedom will certainly improve
a parametrisation of the data.

\subsection{Generalisation to power laws with complex exponents:
log-periodicity}

The idea is to generalise the real exponent $z$ to a complex exponent
$\beta + i \omega$, such that a power law is changed into
$(t_c-t)^{\beta + i \omega}$, whose real part is
$(t_c-t)^{\beta} \cos\left(\omega \ln (t_c-t)\right)$ \cite{reviewsor}.
The cosine will decorate the average power law behaviour with so-called
log-periodic oscillations, the name steming from the fact the oscillations
are periodic in $\ln(t_c -t)$ and not in $t$. As we shall see,
these log-periodic
oscillations can account for a large part of the observed variability around
the power law. Thus, taking them into account provides a better parametrisation
of the data and hence better constraints on the parameters of the power law
$\beta$ and $t_c$.

There are fundamental reasons for introducing log-periodic corrections.
Singularities often exhibit genuine log-periodic corrections that result from
specific mechanisms \cite{reviewsor}:
singularities in the Euler equations with complex exponents have been found
to result from a cascade of Rayleigh-Taylor instabilities leading to
log-periodic oscillatory structures around singular vortices organised
according to discrete self-similar pancakes\cite{Pumiersiggia}; in the
process of formation of black holes, the matter field solution oscillates
periodically in the logarithm of the difference between time and time of
the formation of the singularity \cite{Choptuik};
the phase separation kinetics of a binary
mixture subjected to an uniform shear flow quenched from a disordered to a
homogeneous
ordered phase exhibits log-periodic oscillations due to a cyclical
mechanism of stretching
and break-up of domains, which allows to store and dissipate elastic energy
in the system
\cite{Corberi2}; material failure occurs
after intermittent damage acceleration and quiescent phases that are
well-described by log-periodic structures decorating an overall power law
singularity \cite{failure}; stock market crashes preceded by
speculative bubbles \cite{crash,nasdaq} provide an highly relevant analogy to
the question of sustainability in the growth rate of the human population.
More generally, from the point of view of field theory as a tool-box for
constructing theories of complex systems, we should expect generically the
existence of complex exponents and their associated log-periodic
corrections \cite{salsor}. We suggest that the
presence of log-periodic oscillations deriving from general theoretical
considerations can provide a first step to account for the ubiquitous
observation of cycles in
population dynamics and in economics.

\subsection{Log-periodic fit of the World population}

\subsubsection{Results}

Guided by the recent progress in the understanding of complex systems and
the possibility of complex exponents discussed in the previous section,
we have also fitted the world population data with the following equation
\be \label{lppow}
p(t) \approx A_1 + B_1 (t_c-t)^{\beta}
+C_1(t_c-t)^{\beta}\cos\lp\omega \ln(t_c-t)+\phi \rp~,
\ee
as shown in figure \ref{powlpfits}. We obtained two solutions, the best
having $A\approx 0$,
$B\approx 1624$, $C\approx -127$, $z\approx -1.4$, $t_c \approx 2056$, $\omega
\approx 6.3$ and $\phi \approx 5.1$. The second solution has $A\approx 0.25$,
$B\approx 1624$, $C\approx -127$, $z \approx -1.7$,
$t_c \approx 2079$, $\omega \approx 6.9$ and $\phi \approx -4.4$.
In this extension of equation (\ref{eq:solution}), the cosine term embodies a
discrete scale invariance \cite{Dubrulle} decorating the overall
acceleration with a geometrical scaling ratio $\lambda = \exp \left( 2\pi /
\omega \right)$: the local maxima of the oscillations are converging to
$t_c$ with the
geometrical ratio $1/\lambda$.

\subsubsection{Sensitivity analysis}

Due to the small number of points in the population data set, the
robustness of the fit with equation (\ref{lppow}) was investigated with 
respect to fluctuations in the important physical parameters $t_c$, $\beta$ 
and $\omega$ \cite{comment}. The method we used was as follows. Together 
with the data set (data set 1) obtained
from the United Nations Population Division
Department of Economic and Social Affairs (see the introduction section),
which covers the period $\left[ 0:1998\right]$, seven other
data sets where analysed in an identical manner. These first three data
sets were
generated by removing the first point (data set 2), the two first points
(data set 3) and the 3 first points (data set 4). Hence, those three data
sets cover the periods $\left[ 1000:1998\right]$, $\left[1250:1998\right]$
and $\left[ 1500:1998\right]$. A fifth data set (data set 5) was constructed by
including the UN estimate that the world's population would reach 6 billion in
October 1999 to the original data set (data set 1). Three additional data sets
were created by removing points in the other end from the original data set
(data set 1), {\it i.e}, by removing the last point (data set 6), the two
last points (data set 7) and the three last points (data set 8).  Hence, those
three data sets cover the periods $\left[0:1990\right]$, $\left[0:1980\right]$
and $\left[ 0:1970\right]$.

The differences between the results obtained for the first five data sets
are minor,
as can be seen in Table \ref{tablepop} showing the values corresponding to the
best fits. Data set 6 and 7 are also compatible with the previous 5 whereas
the fit to data set 8 exhibit a significant discrepancy. For $t_c$, this gives
the window
2052 $\pm$ 10 years, which is rather well-constrained.
Furthermore, the values obtained for $\omega \approx 6 \pm 0.5$
(again except for data set 8) are also quite compatible with previous results.
The corresponding fluctuations in the fundamental parameters $z \approx 1.35
\pm 0.11$ and $\lambda \approx 2.8 \pm 0.3$ are also within reasonable bounds.
Note that it is difficult to obtain a better
resolution in time as world population statistics in past centuries
are all generated by using some sort of statistical regression model. This
might explain the relatively low value of the spectral peak obtained
for data set 5, see below.
Furthermore, the peak clearly stands out against the background for seven
out of eight spectra as we now discuss. Another encouraging observation is the
notable amplitude of
the log-periodic oscillations quantified by $C$, approximately $5-10\%$ of
the pure power law acceleration quantified by $B$, as seen in the caption
of figure
\ref{powlpfits}.

\subsubsection{Non-parametric tests of log-periodicity}

We also present a non-parametric test for the existence of the log-periodic
oscillations decorating the spontaneous
singularity, obtained by eliminating the leading trend using the transformation
\be \label{residue}
p\lp t\rp \rightarrow \frac{p\lp t\rp -  A_1 -
B_1(t_c-t)^{\beta}}{C_1(t_c-t)^{\beta}}~ .
\ee
This transformation should produce a pure $\cos\lp \omega \ln(t_c-t)+\phi\rp$
{\it if} equation (\ref{lppow}) was a perfect description. In figure
\ref{reslombpop}, we show the residual defined by (\ref{residue})
for data 3 and data 5 as a function of $\ln (t_c - t)$ as well as
their Lomb periodograms which provide a power spectrum analysis for
unevenly sampled data: the approximately regular oscillations in
$\ln (t_c - t)$ give a significant spectral peak at a  log-angular frequency
$\omega \approx 5.8-6.1$ compatible with the fit of equation (\ref{lppow}),
see Table \ref{tablepop}.

\subsection{Summary}

To sum up the evidence obtained so far, the comparison between the
semi-logarithmic
plots in figures \ref{semilogpop}-\ref{semilogwindex} and the log-log plots
in figures
\ref{pop2030}-\ref{windex2060} validate the power law model (\ref{pow}) at the
expense of the exponential model: there is no doubt that the world
population and major economic and financial indices on average have grown much
faster than
exponentially. The second message is that the rather large fluctuations
decorating
an average power law acceleration can be remarkably well described by a simple
generalisation of the power law in terms of a complex exponent:
not only do we see a good agreement between the spectral analysis
and the fits with equation (\ref{lppow}), in addition
the small fluctuations
in the values for $t_c$, $\beta$ and $\omega$ for the 7 of the 8 data sets
make
the analysis credible for the world population. Of course,
this does not prove that equation (\ref{lppow})
is the correct description and equation (\ref{eq:solution}) is a wrong
description.
However, since the r.m.s. of the fits with the two equations differs by a
factor
of $\approx 4$, there is no doubt that equation (\ref{lppow}) does a better
job
of parameterising the data. This is the numerical argument. The theoretical
justification has already been given above. The two combined certainly makes
the case stronger. For the financial indices, the use of equation
(\ref{lppow})
does not lead to a significant improvement, and this leads us to examine
the relevance of the next order
of the expansion of corrections to the power law.

\section{To second order}

\subsection{Next order of the log-periodic expansion}

The data set containing the Dow Jones Average consists of $\approx 2500$
monthly
quotes for the period $\left[ 1790:1999.9 \right]$.
We propose that it is representative of the capitalistic growth of the
U.S.A. The time span and the sampling rate of this data set makes it reasonable
to use the generalisation (\ref{2feq}) of (\ref{lppow}) to second order
which allows
for a continuous shift in the angular log-frequency $\omega$ \cite{SJ97} in
what effectively corresponds to a Landau or renormalisation group expansion
depending on the prefered framework.

We briefly summarize the method.
Using the renormalization group (RG) formalism on a financial index $I$
amounts
to assuming that the index at a given time $t$ is related to that at
another time $t'$ by the transformations
\be
x' = \phi(x)  ~ ,
\label{firstt}
\ee
\be
F(x) = g(x) + \frac{1}{\mu} F\biggl(\phi(x)\biggl)~,
\label{secondd}
\ee
where $x=t_c-t$.  $t_c$ is the critical time and $\phi$ is called
the RG flow map. Here,
$$
F(x)=I(t_c)-I(t)~,
$$
such that $F=0$ at the critical point and $\mu$ is a constant describing
the scaling of the index evolution upon a rescaling of time (\ref{firstt}).
The function $g(x)$ represents the non-singular part of the function
$F(x)$. We assume as usual that the function $F(x)$ is
continuous and that $\phi(x)$ is differentiable.
In order to use this formalism to constrain the possible time dependence of
the index, we notice that the solution in terms of a power law
of the RG equation (\ref{secondd}) together with  (\ref{firstt}) and the
linear approximation $\phi(x) = \lambda x$ valid close to the critical point
can be rewritten as
\be
{d F(x) \over d\ln x} = \alpha F(x) ~.
\label{ertf}
\ee
This states simply that a power law is nothing but a linear relationship when
expressed in the variables $\ln F(x)$ and $\ln x$. A critical point is
characterized by observables which have an invariant description with
respect to scale transformations on $x$. We can exploit this and
the expression (\ref{ertf}) to propose the structure
of the leading corrections to the power law with log-periodicity. Hence, we
notice
that (\ref{ertf}) can be interpreted as a bifurcation equation for the
variable $F$
as a function of a fictitious ``time'' ($\ln x$) as a function of the
``control
parameter'' $\alpha$. When $\alpha > 0$, $F(x)$ increases with $\ln x$
while it
decreases for $\alpha < 0$. The special value $\alpha = 0$ separating the
two regimes
corresponds to a bifurcation. Once we have recognized the
structure of the expression (\ref{ertf}) in terms of a bifurcation, we can
use the
general reduction theorem telling us that the structure of the
equation for $F$ close to the bifurcation can only take a universal {\it
non-linear} form given by
\be
{dF(x) \over d\ln x} =  (\alpha + i \omega) F(x)
+ (\eta + i \kappa) |F(x)|^2 + {\cal O}(F^3)   .
\label{azepo}
\ee
where $\alpha>0$, $\omega$, $\eta$ and $\kappa$ are real coefficients and
${\cal O}(F^3)$ means that higher order terms are neglected. The
generality of this expression stems from the fact that it is nothing but a
Taylor's expansion of a general functional form ${dF(x) \over d\ln x} =
{\cal
F}(F(x))$. Such expansions are known in the physics literature as Landau
expansions. We stress that this expression represents a non-trivial
addition to the theory, constrained uniquely by symmetry laws.
Going up to second order included, equation $(\ref{lppow})$ becomes \cite{SJ97}
\be
\ln\lp I\lp t \rp \rp = A_2 + B_2 {\lp t_c- t \rp ^{\beta} \over \sqrt{ 1 +
\lp { t_c- t \over \tau} \rp^{2 \beta}}} \left[ 1+ C_2 \cos \biggl(
\omega \ln \lp t_c-t \rp + {\Delta \omega \over 2 \beta} \ln \lp 1 +
\lp {t_c- t \over \tau} \rp^{2\beta} - \phi \rp \biggl) \right]~.
\label{2feq}
\ee
This extension has been found useful in order to account for the behaviour of
stock market prices before large crashes over extended period of times up to
8 years. The present analysis thus constitutes a major generalisation as it
includes over 200 years of data. Previous work have established
a robust and universal signature preceding large crashes occuring in major
financial stock markets, namely accelerated price increase decorated by large
scale log-periodic oscillations culminating in a spontaneous singularity
(critical point). The previously reported cases, which are well-described by
equation $(\ref{lppow})$, comprise the Oct. 1929 US crash, the Oct. 1987 world
market crash, the Jan. 1994 and Oct. 1997 Hong-Kong crashes, the Aug.~1998
global market event, the April 2000 Nasdaq crash,
the 1985 Forex event on the US dollar, the correction on the US dollar
against the Canadian dollar and the Japanese Yen starting in Aug.~1998, as
well as the bubble on the Russian market and its ensuing collapse in June 1997
\cite{crash,nasdaq}. Furthermore, twenty-one significant bubbles followed
by large
crashes or by severe corrections in the stock markets indices of the South
American and Asian countries, which exhibit log-periodic
signatures decorating an average power law acceleration,
have also been identified \cite{emerg}. In all these
analyses, the time scales have been restricted to 1 to 8 years. In
contrast, the general renormalisation group theory of such spontaneous
singularities allow for an hierarchy of critical points at all scales
\cite{erzan,SSS}. The results given below suggest that singularities do indeed
cascade in a robust way up to the largest time scales or conversely from the
largest scale to the smallest scales \cite{Drozdz}.

\subsection{Second order fit of the Dow Jones Average index}

\subsubsection{Methodology}

We fit the logarithm of the extended Dow Jones index to equation (\ref{2feq}).
As mentioned,
taking the logarithm provides in our opinion the simplest and most
robust way to account for inflation. Furthermore, taking the
logarithm embodies the notion that only relative changes are important.
Another more subtle reason can be given in terms of the magnitude of the
crash following the singularity: a simple model of rational expectations
\cite{bubmodel}
shows that if the loss during a crash is proportional to the maximum price,
then the relevant quantity is the logarithm of the price in accordance
with the standard economic notion that only relative changes should be
relevant.

Fitting equation (\ref{2feq}) to some data set is difficult even with a large
data set (for noisy data with only a few hundred points or less, it becomes
quite
impossible), due to the degenerate r.m.s. landscape corresponding to the
existence of many local minima as a function of the free parameters $t_c$,
$\beta$, $\omega$, $\tau$ $\Delta \omega$ and $\phi$. This means that the
r.m.s. alone is not a good measure of the quality of the fit and additional
{\em physical} constraint are needed as discriminators. This has been
discussed
at length in \cite{JLS}. In brief, we will demand that the value of
$\beta$ and $\omega$ are compatible with what has been found previously for
large crashes and that the value of the transition time $\tau$ between the
two competing frequencies is compatible with the time window $t_c - t_0$,
where $t_0$ is the date of the first data point and $t_c$ the date of the
singularity in the first derivative. Unfortunately, we have no means
to impose a criterion on the frequency-shift $\Delta \omega$.

 Specifically,
we will demand that $0.2 < \beta < 0.7$, $4.5 < \omega < 9$ and $\lp t_c -
t_0 \rp/3 < \tau \stackrel{<}{\sim} \lp t_c - t_0 \rp$ and the more the
parameters fall in the mid-range, the higher confidence is attributed to the
fit. These constraints are similar to what was used in \cite{JLS} except for
the constrain on $\tau$ which upper limit has been made stricter here. The
reason
for this is simply that, whereas in the cases of the 1929 and 1987 stock market
crashes on Wall Street, it was not obvious to decide the starting date of the
bubble, it is now objectively determined by a historical event being the
creation of the U.S.A. as an independent nation \cite{UShistory}.
The parameter values of the
five qualifying fits is shown in Table \ref{tabledj}. We stress that the
majority
of the fits were discarded due to rather large values for either $\omega$ or
$\tau$ or negative values for $\omega$. We see that the best fit in terms of
the r.m.s. also has the most reasonable parameter values for $\beta$, $\omega$
and $\tau$ in terms of the discussion above.

\subsubsection{Results}

\begin{figure}
\begin{center}
\epsfig{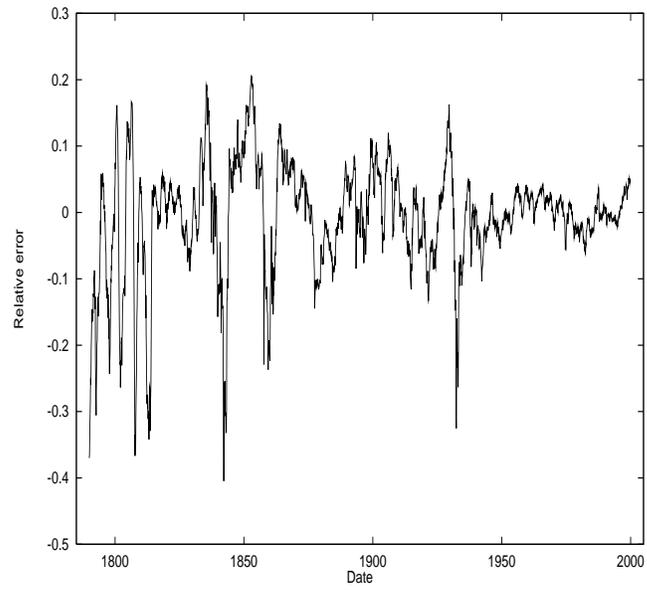}
\caption{\protect\label{reslombdj} Relative error between the fit
with equation $(\protect\ref{2feq})$ and the data as a function of time.}
\vspace{12cm}
\end{center}
\end{figure}

The best fit of equation (\ref{2feq})) to the $210$ years of monthly quotes is
shown in figure \ref{djfit} and its parameter values are given in the caption.
Note that the value of the angular log-frequency $\omega \approx 6.5$ compared
to $\omega \approx 6.3$ as well as the value for the position of the
singularity
$t_c \approx 2053$ compared $t_c \approx 2056$ are in close agreement with the
values found for the analysis of the world population. Furthermore,
the cross-over time scale $\tau \approx 171$ years is perfectly compatible with
the total time window of 210 years. In figure \ref{reslombdj}, the relative
error
between the fit and the data is shown. We see that the error fluctuates nicely
around zero as it should. Furthermore, the error is decreasing from left
to right clearly showing that the acceleration in the data is better
and better modeled by equation (\ref{2feq})) as we approach the present.
This behaviour is in fact to be expected from an equation such as
(\ref{aafajdak}) allowing for an additive noise term to describe other sources
of uncertainties: using
the Fokker-Planck formalism, one can show that, as the singularity at $t_c$ is
approached, the noise term becomes negligible and the acceleration of the data
should approach better and better a pure power law. This can also be seen
directly from (\ref{aafajdak}) with an additive noise: the
divergence of $p(t)$ dwarves any bound noise contribution.

\subsubsection{Discussion}

The inset of figure \ref{djfit} shows the extrapolation of the
fit up to the critical time $t_c=2053$. It suggests that the Dow Jones
index will
climb to impressive values in the coming decades from its present level
around 11,000 at the beginning of year 2000.
It is interesting that this resonates with a series of claims that the Dow
Jones
will climb to
36,000 \cite{Glassman}, 40,000 \cite{Elias} or even 100,000 \cite{Kadlec}
 in the next two or three decades.
Glassman, an investing columnist for
the Washington Post, and Hassett, a former senior economist with the Federal
Reserve, develop the argument that stocks have been
undervalued for decades and that, for the next few years, investors can
expect a
dramatic one-time upward adjustment in stock prices \cite{Glassman}.
Elias, a financial advisor and author, believes that
forces such as direct foreign investment, domestic savings, and cooperative
central-banking policies will drive the vigorous market, as will the
dynamics of
the New Economy, which allows for the coexistence of high economic growth, low
interest rates, and low inflation. In his view, the Dow Jones could reach
40,000
around 2016 \cite{Elias}. Kadlec, chief investment strategist for Seligman
Advisors Inc.
predicts that the Dow
Jones Industrial Average will end up at 100,000 in the year 2020
\cite{Kadlec}.
We find that equation (\ref{2feq}) predicts that the level 36,000-40,000
will be reached in 2018-2020 A.D. and the level 100,000 in 2026 A.D, not far
from these claims! Of course, the extrapolation of this growth
closer to the singularity becomes unreliable due to standard limitations, such
as finite size effects, and must be taken with a ``hand-full of salt''.

In the academic financial literature, a time series such as the Dow Jones
shown in figure \ref{djfit} has been argued to exhibit an anomalously large
return, averaging $6\%$ per year over the 1889-1978 period \cite{Mehra},
which cannot be explained by any reasonable risk aversion coefficient.
A solution for this puzzle is that infrequent large crashes occur or even a
major still untriggered crash is looming over us; in this interpretation,
the ``anomalous'' return becomes the normal remuneration for the risk to
stay invested in the market \cite{Ritz}. Our analysis suggests that the
situation
is even worse than this: not only the market has a large growth rate but
this growth rate is {\it accelerating} such that the market is growing as
a power law towards a spontaneous singularity.

\section{Synthesis and theoretical discussion}

\subsection{Summary}

The fact, that both the human world population over two thousand years, the
GDP of the world and six national, regional and world financial indices over
most of their lifespan agree both in i) the prediction of a spontaneous
singularity, ii) the approximate location of the critical time and iii) the
approximate self-similar
patterns decorating the singularity is quite remarkable to say the least. This
suggests that they may have a closely correlated dynamics, in fact more than
the coupling between population $p(t)$ and carrying capacity $K(t)$ written
in equations such as $(\ref{ajdak})$ would make us believe. The outstanding
scientific question is whether the rate of innovations fueling the economic
growth is a random process on which industrial and population selection
operates or if it is driven by the pressing needs of the growing population.
The main message of this study is that, whatever the answer and irrespective
of one's optimistic or pessimistic view of the world sustainability, these
important pieces of data all point to the existence of an end to the
present era, which will be irreversible and cannot be overcome by any novel
innovation of the preceding kind, {\it e.g.}, a new technology that makes
the final conquest of the Oceans and the vast mineral resources there
possible. This, since any new innovation is deeply  embedded in the very
existence of a singularity, in fact it {\it feeds} it. As a result, a
future transition of mankind towards a qualitatively new level is quite
possible.

The reader not familiar with critical phenomena and singularities
\cite{DombGreen,critical,Dubrulle}
may dismiss our approach without further ado on the basis that
all demographic insights show that the population
growth is now decelerating rather than accelerating. 
Indeed, many developed countries show a substantial reduction in fertility.
However, ``the tree should not hide the forest'' as the proverb says, in other words
this deceleration is compatible with the concept of a finite-time 
singularity in the presence of so-called ``finite size effects''
\cite{finitesizeeffect}.
Namely, it is well-known that nature does not have pure singularities in the
mathematical sense of the term. Such critical points are
always rounded off or
smoothed out by the existence of friction and dissipation and by the finiteness
of the system. This is a well-known feature of critical points 
\cite{finitesizeeffect}. Finite-time singularities are similarly rounded-off
by frictional effects, A clear example is provided by Euler's disk
\cite{Moffatt}, a rotating coin settling to rest in finite time after, in 
principle, an infinite number of rotations. In reality, the rotational speed
accelerates until a point when friction due to air drag and solid contact with the
support saturate this acceleration and stop the rotation abruptly. 
The upshot here is that finite-size effect and friction do not prevent the 
effect we document here to be present, namely the acceleration of the growth rate,
up to a point where the proximity to the critical point makes finite size
effects and dissipation-like effects to take over. The fact that these
``imperfections'' become relevant in the ultimate stage of the trajectory
does not change the validity of the conclusions. The change of regime to a new
phase subsists. Only its absolute abruptness is replaced by a somewhat
smoother transition, albeit still rather sharp on the time scale of the 
total time span. In the present context,
the observed very recent deceleration of the growth rate can be taken as
a signature that mankind is entering in the critical region towards a 
transition to a new regime. Since the world population growth rate topped
in 1970, this corresponds to approximately 80 years from the predicted
critical point, or only $4\%$ of the total 
timespan of the investigated time series.

\subsection{Related work}

Other authors have documented a super-exponential acceleration of human
activity.
Kapitza has recently analysed the dynamical evolution of the human
population \cite{Kapitza},
both aggregated and regionally and also documents a consistent overall
acceleration
until recent times. He introduces a saturation effect to limit the blow-up and
discuss different scenarios. Using data from the Cambridge encyclopedia,
he argues that epochs of characteristic evolutions
or changes shrink as a geometrical series. In other words, the epoch sizes
are approximately
equidistant in the logarithm of the time to present.
In a study of an important human activity,
van Raan has found that the scientific production since the 16th century
in Europe has accelerated much faster than
exponentially \cite{Raan}.
Using the data of DeLong \cite{Bradford},
Hanson finds that the history of the world economic production since
prehistoric times
can only be accounted for by adding
three exponentials, each one being interpreted as a new ``revolution''
\cite{Hanson}:
hunting followed by farming and then by industry. He finds that each
exponential mode grew over one hundred times
faster than its predecessor. He also plots the logarithm of the
world product as a function of the logarithm of $t_c-t$ with $t_c =2050$
and find a
reasonable straight line decorated by oscillations marking the different
transitions.

Macro-economic models have been developed that predict the possibility of
accelerated growth
\cite{Romer}. Maybe the simplest model is that of Kremer \cite{Kremer} who
notes that,
over almost all human history, technological progress has led mainly to
an increase in
population rather than an increase in output per person. In his model,
the economic output
per person $Y(t)/L(t)$, where $Y(t)$ is the total output comprising all
artifacts
 and $L(t)$ is the
total population,
is thus set equal to the subsistence level ${\bar y}$ which is assumed fixed:
\be
{Y(t) \over L(t)} = {\bar y}~. \label{fjak}
\ee
The output is supposed to depend on technology and knowledge
$A(t)$ and labour (proportional to $L(t)$):
\be
Y(t) = Y_0 \left[ A(t) L(t) \right]^{1-\alpha}~, \label{fjaka}
\ee
where $0 < \alpha < 1$. The growth rate of knowledge and technology is taken
proportional to population and to knowledge:
\be
{dA \over dt} = B L(t) A(t)~, \label{fbnkala}
\ee
embodying the concept that a larger population offers more opportunities
for finding
exceptionally talented-people who will make important innovations and that
new knowledge
is obtained by leveraging existing knowledge.
Eliminating $Y(t)$ and $A(t)$ between (\ref{fjak}-\ref{fbnkala}) gives the
equation for the
total population:
\be
{dL \over dt} = {1-\alpha \over \alpha} B~ [L(t)]^2~.   \label{fakal}
\ee
This is the case $\delta = 1$  of equation (\ref{aafajdak}), showing that
the population
and its output develop a finite-time singularity (\ref{pow}) with the exponent
$z=-1$. Kremer tested this prediction by using population estimates extending
back to 1 million B.C., constructed by archaeologists and anthropologists:
he showed
that the population growth rate is approximately linearly increasing with
the population
\cite{Kremer},
in agreement with (\ref{fakal}). Our result $z\approx -1.9$ for the human
population
exaggerates the singularity. On the other hand, as shown in Table 1, we find
a remarkable consistent value $z \approx -1$ for all financial indices. Our
refinements with the log-periodic formulas in order to account for the
significant
structures decorating the average power laws necessary lead to deviations
from this
``mean-field'' value, which should be considered as an approximation
neglecting the effect
of fluctuations.
This theory also predicts, in agreement with historical
facts, that in the historical times when regions were separated,
 technological progress was faster in regions with larger population, thus
explaining
 the differences between Eurasia-Africa, the Americas, Australia and Tasmania.

\subsection{Multivariate finite-time singularities}

Kremer's model is only one of a general class of growth models \cite{Romer}.
We briefly recall the general framework developed by Romer \cite{Romer2},
which allows us to generalise the concept of
finite-time singularities to multivariate dynamics and to
exhibit the structure of its solution and follow \cite{Romer} in our
exposition.
The model involves four variables,
labour $L$, capital $K$,
technology $A$ and output $Y$. There are two sectors, a goods-producing
sector where
output is produced and an R\&D sector where additions to the stock of
knowledge are made.
The fraction $a_L$ of the labour force is used in the R\&D sector and the
fraction $1-a_L$  in
the goods-producing sector; similarly, the fraction $a_K$ of the capital
stock is used in R\&D
and the rest in goods production.  Both sectors use the full stock of
knowledge. The quantity
of output produced at time $t$ is defined as
\be
Y(t) = \left[ (1-a_K) K(t) \right]^{\alpha}~ \left[A(t)(1-a_L)
L(t)\right]^{1-\alpha}~,
\label{fjakala}
\ee
with $0 < \alpha < 1$. Expression (\ref{fjakala}) uses the so-called
Cobb-Douglas functional
form with power law relationships which imply constant returns to capital
and labour: within
a given technology, doubling the inputs doubles the amount that can be
produced. Expression (\ref{fjakala}) writes that the economic output increases
with invested capital, with technology and R\&D and with labor.

The production of innovation is written as
\be
{dA \over dt} = B \left[ a_K K(t)\right]^{\beta}~\left[ a_L
L(t)\right]^{\gamma}~
\left[A(t)\right]^{\theta}~,~~~~~B>0,~~\beta \geq 0, ~~ \gamma \geq 0~.
\label{fjaklaaa}
\ee
The growth of knowledge is thus controlled by the pre-existing knowledge,
by capital
investment in research and by the size of the population of innovators.

As in the Solow model \cite{Romer}, the saving rate $s$ is exogenous and
constant and depreciation
is set to zero for simplicity so that
\be
{dK \over dt} = s Y(t) = s \left[ (1-a_K) K(t) \right]^{\alpha}~
\left[A(t)(1-a_L) L(t)\right]^{1-\alpha}~.  \label{fqqoloa}
\ee

Let us consider (\ref{fjaklaaa}). If $K$ and $L$ are constant, it reduces
to an equation of
the form (\ref{aafajdak}), which exhibits a finite-time singularity only
for $\theta > 1$.
In the presence of the coupling to the other growing dynamical variables
$K$ and $L$,
a finite-time singularity may occur even in the situation $\theta <1$.

As a first example, let us consider the case of a fixed population $L(t)
=$ constant.
Equations (\ref{fjaklaaa}) and (\ref{fqqoloa}) can be rewritten as
\bea
{dA \over dt} &=& b A^{\theta} K^{\beta}~. \label{nfncbbzb}\\
{dK \over dt} &=& a A^{1-\alpha} K^{\alpha} ~,  \label{vbxz}
\eea
We look for the condition on the exponents such that $A(t)$ and $K(t)$ exhibit
a finite-time singularity. We thus look for solutions of the form
\bea
A(t) &=& A_0 (t_c-t)^{-\delta}~, \label{bcvccz}\\
K(t) &=& K_0 (t_c-t)^{-\kappa}~,  \label{nbeq}
\eea
with $\delta$ and $\kappa$ positive.
Inserting these expressions in (\ref{vbxz}) and (\ref{nfncbbzb}) leads to
two equations for the two exponents $\delta$ and $\kappa$ obtained from the
conditions that
the powers of $(t_c-t)$ are the same on the r.h.s. and l.h.s. of
(\ref{vbxz}) and (\ref{nfncbbzb}).
Their solution is
\bea
\delta &=& {1+\beta - \alpha \over (1-\alpha)(\theta+\beta-1)}~,\\
\kappa &=& {2-\theta - \alpha \over (1-\alpha)(\theta+\beta-1)}~.
\eea
The condition that both $\delta$ and $\kappa$ are positive enforce that
$\theta+\beta > 1$, which is the condition replacing $\theta >1$ for
the existence of a finite-time singularity in the monovariate case.
This shows that the combined effect of past innovation and capital has
the possibility of creating an explosive growth rate even when {\it each}
of these factors in isolation does not. Note that inequality $\theta+\beta
> 1$ ensures that $\delta > \kappa$, {\it i.e.}, the growth of the
technological stock is faster than that of the capital.

There are many ways to reinsert the dynamical evolution of the population.
Let us here
consider the simplest one used by Kremer \cite{Kremer}, which consists in
assuming that
$L(t)$ is proportional to $K(t)$ as given by (\ref{fjak}). Then,
expressions (\ref{fjaklaaa})
and(\ref{fqqoloa}) give
\bea
{dA \over dt} &=& a' \left[L(t) \right]^{\beta +
\gamma}~\left[A(t)\right]^{\theta}~,
~~~~~a'>0,~~\beta \geq 0, ~~ \gamma \geq 0~,   \label{fjakelaawa}  \\
{dL \over dt} &=& b'  L(t) \left[A(t) \right]^{1-\alpha}~.  \label{fqasfqoloa}
\eea
Looking for solutions of the form (\ref{bcvccz}) and (\ref{nbeq}) gives
\bea
\delta &=& {1 \over 1-\alpha}~,\\
\kappa &=& {2-\theta - \alpha \over \beta + \gamma}~.
\eea
It is interesting to find that the technology growth exponent $\delta$ is
not at all controlled by $\theta$ nor $\beta$ and $\gamma$. This
illustrates that a finite-time singularities can be created
from the interplay of several growing variables resulting in a non-trivial
behaviour. In the present context, it means that the interplay between
different quantities, such as capital and technology, may produce an
``explosion'' in the population even though the individual dynamics
do not. In particular, this interplay provides an explanation of our finding
of the same value of the critical time $t_c (\approx 2052 \pm 10)$ both for
the
population and economic indices.

\section{Possible scenarios}

We now attempt to guess what could be the possible scenarios for mankind
close to and beyond the critical time $t_c$.

A gloomy scenario is that humanity will enter a
severe recession fed by the slow death of its host (the Earth), in the
spirit of
the analogy \cite{Hern} proposed between the human species and cancer.
This worry about human population size and growth is shared by many
scientists, including
the Union of Concerned Scientists (comprising 99 Nobel Prize winners) which
asks
nations to
``stabilise population.'' Representatives of national
academies of science from throughout the world met in New Delhi, 24-27 October
1993, at a ``Science Summit'' on World Population. The participants issued a
statement, signed by representatives of 58 academies on population issues
related to
development, notably on the determinants of fertility and concerning the effect
of demographic growth on the environment and the quality of life.
The statement finds that ``continuing population growth poses a great
risk to humanity,'' and proposes a demographic goal: ``In our judgment,
humanity's ability to
deal successfully with its social, economic, and environmental problems will
require the achievement of zero population growth within the lifetime of our
children'' and  ``Humanity is
approaching a crisis point with respect to the interlocking issues of
population,
environment and development because the Earth is finite''
\cite{statementAcads}.
Possible scenarios involve a systematic development of terrorism and the
segregation of
mankind into at least two groups, a minority of wealthy communities hiding
behind
fortresses from the crowd of ``barbarians'' roaming outside, as discussed
in a recent seminar at the US National Academy of Sciences. Such a scenario
is also quite possible for the relation between developed and developing
countries.

On a more positive note, it may be that ``ecological'' actions of the kind
mentioned above will grow
in the next decades, leading to a smooth transition towards an
ecologically-integrated
industry and humanity. Some signs may give indications of this path: during the
1990s, wind power has been growing at a rate of 26\% a year and solar
photo-voltaic power
at 17\% compared to the growth in coal and oil under 2\%; governments have
``ratified'' more
than 170 international environmental treaties, on everything from fishing
to decertification
\cite{stateofworld}. However, there are serious resistances \cite{Nego},
in particular because there is no consensus on the seriousness of the
situation:
for instance, the economist J.L. Simon writes that ``almost every
measure of material and environmental human welfare in the United States
and in the World
shows improvement rather than deterioration'' \cite{Simon}.
It may be that the strikingly
similar explosive trend in population and GDP would not necessarily
persist in the future when taking the differences between
regional developments into account. Perhaps what is needed to avoid the
finite-time
singularity is a massive transfer of resources from developed to developing
countries. The recent discussions at the G7/8 summit indicates that
the developed world is becoming increasingly aware of the discrepancy.

Extrapolating further, the evolution from a growth regime to a balanced
symbiosis with nature
and with the Earth's resources requires the transition to a
knowledge-based society, in which
knowledge, intellectual, artistic and humanistic values replace the quest
for material wealth.
Indeed, the main economic difference is that ``knowledge'' is non-rival
\cite{Romer2}:
the use of an idea or of a piece of knowledge in one place does not prevent
it from being
used elsewhere; in contrast, say an item of clothing by an individual
precludes its
simultaneous use by someone else. Only the emphasis on non-rival goods will
limit
ultimately the plunder of the planet. Some so-called
``primitive'' societies seem to have been able to evolve
into such a state \cite{Gunsdiam}.

The race for growth could however continue or even be enhanced if fundamentally
new discoveries at a different level of the hierarchy witnessed until present
enabled mankind to start the colonisation of other planets. The conditions for
this are rather drastic, since novel modes of much faster propulsions are
required as
well as revolutions in our control of the adverse biological effects of
space on
humans. It may be that some evolved form of humans
will appear who are more adapted to the hardship of space.
This could lead to a new era of renewed accelerated growth after a period of
consolidation, culminating in a new finite-time singularity, probably
centuries in the future.

{\bf Acknowledgement:} We thank P. Kendall and R. Prechter for help in
providing the
financial data from the Foundation For The Study Of Cycles, R. Hanson for
the world GDP data and useful discussions,
B. Taylor of Global Financial Data for the permission to use their data,
M. Lagier, D. Zajdenweber for discussions, U. Frisch and D. Stauffer for a
critical reading of
the manuscript and for useful suggestions.

\vskip 0.5cm
{Note Added in Proofs}:
Nottale, Chaline and Grou \cite{Chaline2,Chaline3} have recently 
independently applied a log-periodic analysis to the main crises of different
civilisation. They first noticed that historical events seem to accelerate. This was
actually anticipate by Meyer who used a primitive for of 
log-periodic acceleration analysis \cite{Meyer1,Meyer2}. Grou \cite{Grou} has
demonstrated that the economic evolution since the neolithic can be described in
terms of various dominating poles which are subjected to an accelerating crisis/
no-crisis pattern. Their quantitative analysis on the
median dates of the main periods of economic crisis in the history of Western
civilization (as listed in \cite{Grou,Braudel,Gilles} are as follows (the dominating pole
and the date are given in years / JC): \{Neolithic: -6500\}, \{Egypt: -3000\},\{Egypt: -900\},
\{Grece: -100\}, \{Rome: +400\}, \{Byzance: +800\}, \{Arab expansion: +1100\}, \{Southern
Europ: +1400\}, \{Netherland:+1650\}, \{Great-Britain: +1775\}, \{Great-Britain: +1830\},
\{Great-Britain: +1880\}, \{Great-Britain: +1935\}, \{United-States: +1975\}.
Log-periodic acceleration with scale factor $\lambda = 1.32 \pm 0.018$ occurs towards 
$t_c = 2080 \pm 30$. Agreement between the data and the log-periodic law is
statistically highly significant ($t_{\rm student}  = 145$, Proba $<< 10^{-4}$).
It is striking that this independent analysis based on a different data set gives 
a critical time which is compatible with our own estimate $2052 \pm 10$.

\begin{figure}
\begin{center}
\parbox[l]{8.5cm}{
\epsfig{file=fitlogeafe2060.eps,height=8cm,width=8.5cm}
\caption{\label{eafe2060}The EAFE index as a function of $t_c - t$ with
$t_c=2060$.
The straight line is the
fit with a power law $p(t) = a(t_c - t)^z$ with $t_c=2060$ fixed, see Table
\protect\ref{ztable}.}}
\hspace{5mm}
\parbox[r]{8.5cm}{
\epsfig{file=fitlogwindex2040.eps,height=8cm,width=8.5cm}
\caption{\label{windex2040}The World index as a function of $t_c - t$ with
$t_c=2040$.
The straight line is the
fit with a power law $p(t) = a(t_c - t)^z$ with $t_c=2040$ fixed, see Table
\protect\ref{ztable}.}}

\vspace{1.5cm}

\parbox[l]{8.5cm}{
\epsfig{file=fitlogwindex2050.eps,height=8cm,width=8.5cm}
\caption{\label{windex2050}The World index as a function of $t_c - t$ with
$t_c=2050$.
The straight line is the
fit with a power law $p(t) = a(t_c - t)^z$ with $t_c=2050$ fixed, see Table
\protect\ref{ztable}.}}
\hspace{5mm}
\parbox[r]{8.5cm}{
\epsfig{file=fitlogwindex2060.eps,height=8cm,width=8.5cm}
\caption{\label{windex2060}The World index as a function of $t_c - t$ with
$t_c=2060$.
The straight line is the
fit with a power law $p(t) = a(t_c - t)^z$ with $t_c=2060$ fixed, see Table
\protect\ref{ztable}.}}
\end{center}
\end{figure}


\begin{figure}
\begin{center}
\parbox[l]{8.5cm}{
\epsfig{file=datset3-5.res.eps,height=8cm,width=8.5cm}}
\hspace{5mm}
\parbox[r]{8.5cm}{
\epsfig{file=datset3-5.fp.eps,height=8cm,width=8.5cm}}
\caption{\protect\label{reslombpop}Left: Residue between best fit and data sets
3 and 5, as defined by equation $(\protect\ref{residue})$. Right: Spectrum of
residue using a Lomb periodogram. The position of the peak corresponds to
$\omega \approx 5.8$, which should be compared with $\omega \approx 6.5$ for
the fit with equation $(\protect\ref{lppow})$ for data set 5. For data set 3,
the peak corresponds to $\omega \approx 6.1$, which should be compared with
$\omega \approx 6.5$ for the fit.}

\vspace{0.5cm}

\epsfig{file=FigUSmarket.eps,height=10cm,width=14cm}
\caption{\protect\label{djfit} The circles are the logarithm of the yearly
quotes from Dec. 1790 to Dec. 1999, which are shown instead of the monthly
quotes used in the fit to better show the two fit curves.
The upward trending full line is the best fit with equation
$(\protect\ref{eq:solution})$ with $\mbox{r.m.s.} = 0.307$, $A \approx -13.7$,
$B \approx 70.8$, $z\approx -0.27$ and $t_c \approx 2068$. The full
oscillating line is the best fit with
equation $(\protect\ref{2feq})$ to the extended Dow Jones for the period
$\left[ 1790:1999.75 \right]$. The fit gives $\mbox{r.m.s}=0.236$,
$A\approx 25.1$, $B\approx -4.13$, $C\approx -0.055$, $\beta \approx 0.39$,
$t_c \approx 2053$, $\omega  \approx 6.5$, $\tau \approx 171$, $\Delta \omega
\approx -58$ and $\phi \approx -5.8$. The inset shows the extrapolation of the
fit up to the critical time $t_c \approx 2053$.}
\end{center}
\end{figure}


\mbox{ }
\newpage

\begin{table}
\begin{center}
\begin{tabular}{|c|c|c|} \hline
Index &  Year  & $z$  \\ \hline
DJ    &  2040  & $-0.68$      \\ \hline
DJ    &  2050  & $-0.77$      \\ \hline
DJ    &  2060  & $-0.86$      \\ \hline
S\&P    &  2040  & $-1.10$      \\ \hline
S\&P    &  2050  & $-1.25$      \\ \hline
S\&P    &  2060  & $-1.40$      \\ \hline
Latin Am    &  2040  & $-0.89$      \\ \hline
Latin Am     &  2050  & $-1.04$      \\ \hline
Latin Am     &  2060  & $-1.18$      \\ \hline
Europe    &  2040  & $-0.89$      \\ \hline
Europe    &  2050  & $-1.05$      \\ \hline
Europe    &  2060  & $-1.20$      \\ \hline
EAFE    &  2040  & $-0.87$      \\ \hline
EAFE    &  2050  & $-1.00$      \\ \hline
EAFE    &  2060  & $-1.13$      \\ \hline
World    &  2040  & $-0.88$      \\ \hline
World    &  2050  & $-1.01$      \\ \hline
World    &  2050  & $-1.14$      \\ \hline
\end{tabular}
\end{center}
\caption{\label{ztable}Values for the exponent $z$ from the fits with
equation (\ref{pow}) shown in figures \ref{pop2030} - \ref{windex2060}  }
\end{table}

\begin{table}
\begin{center}
\begin{tabular}{|c|c|c|c|c|c|c|c|c|} \hline
data & number of points & time period & $t_c$    & $\beta$     & $\omega$ &
$\lambda$ & $\omega_{\rm spectrum}$ & Peak power \\ \hline
set 1 & $18$ & $\left[ 0:1998\right]$ & $2056$ & $-1.39$ & $6.3$ & $2.7$ &
$5.7$ & $4.3$
\\ \hline
set 2 & $17$ & $\left[ 1000:1998\right]$ & $2053$ & $-1.35$ & $6.2$ & $2.8$
& $5.8$ & $5.0$
\\ \hline
set 3 & $16$ & $\left[ 1250:1998\right]$ &  $2059$ & $-1.45$ & $6.5$ & $2.6$
& $5.8$ & $5.9$
\\ \hline
set 4 & $15$ & $\left[ 1500:1998\right]$ & $2058$ & $-1.43$ & $6.5$ & $2.6$
&  $5.9$ & $6.1$
\\ \hline
set 5 & $19$ & $\left[ 0:1999.75\right]$ & $2062$ & $-1.46$ & $6.5$ & $2.6$
& $6.1$  & $4.5$ \\ \hline
set 6 & $17$ & $\left[ 0:1990\right]$ & $2043$ & $-1.24$ & $5.8$ & $2.9$
& $5.3$  & $3.4$  \\ \hline
set 7 & $16$ & $\left[ 0:1980\right]$ & $2043$ & $-1.25$ & $5.5$ & $3.1$
& $5.2$  & $3.4$  \\ \hline
set 8 & $15$ & $\left[ 0:1970\right]$ & $2034$ & $-1.20$ & $4.9$ & $3.9$
& $5.1$  & $3.4$   \\ \hline
\end{tabular}
\end{center}
\caption{\label{tablepop} $t_c$ is the critical time predicted
from the fit of the world population data to equation (\ref{lppow}). The
other physical parameters $\beta$ and $\omega$ of the fit are also shown.
$\lambda = \exp \left( 2\pi / \omega \right)$ is the prefered scale
ratio of the underlying dynamics.
$\omega_{\rm spectrum}$ is the angular log-frequency obtained from the
non-parametric spectral analysis of the log-periodic oscillations.}
\end{table}

\begin{table}
\begin{center}
\begin{tabular}{|c|c|c|c|c|c|} \hline
minima & $t_c$    & $\beta$     & $\omega$ & $\tau$ & r.m.s. \\ \hline
first  &  $2053$  & $0.39$ & $6.5$   & $171$ & $0.23582$ \\ \hline
second &  $2046$  & $0.36$ & $5.3$   & $240$ & $0.23584$ \\ \hline
third  &  $2067$  & $0.42$ & $6.8$   & $122$ & $0.23644$ \\ \hline
fourth &  $2009$  & $0.61$ & $5.5$   & $188$ & $0.27459$ \\ \hline
fifth  &  $2007$  & $0.62$ & $4.8$   & $206$ & $0.27461$ \\ \hline
\end{tabular}
\end{center}
\caption{\label{tabledj} $t_c$ is the critical time predicted
from the fit of the logarithm of the extended Dow Jones for the period
$\left[ 1790:1999.75 \right]$ to equation(\ref{2feq}). The other
physical parameters $\beta$, $\omega$  and $\tau$ of the fits are also shown.}
\end{table}


\newpage



\begin{thebibliography}{}


\bibitem{benderorszag} Bender C, Orszag S.A (1978)
page 147 in {\it Advanced Mathematical Methods for Scientists and Engineers.}
McGraw-Hill, New York. 

\bibitem{Bradford} J. Bradford DeLong (1998) 
Estimating World GDP, One Million B.C. - Present. Working paper available at 
http://econ161.berkeley.edu/TCEH/1998\_Draft/World\_GDP/Estimating\_World\_GDP.html

\bibitem{stateofworld} Brown LR, Flavin C (1999) 
{\it State of the World, Millenium edition}. A Worldatch Institute report on 
Progress towards a sustainable society, W.W. Norton \&Co. and Worldwatch 
Institute.

\bibitem{finitesizeeffect} Cardy, J.L. editor (1988)
Finite-size scaling (Amsterdam; New York:
North-Holland; New York, NY, USA; Elsevier Science Pub. Co).

\bibitem{Choptuik} Choptuik M.W (1999) Universality and 
scaling in gravitational collapse of a massless scalar. {\it Physical Review 
Letters} 70: 9-12 \& Critical behaviour in gravitational collapse. {\it 
Progress of Theoretical Physics} Supplement 136: 353-365.

\bibitem{Cohenscience} Cohen J.E. (1995) Population growth and 
Earth's human carrying capacity. {\it Science} 269: 341-346.

\bibitem{Corberi2} Corberi F, Gonnella G, 
Lamura A (2000) Structure and rheology of binary mixtures in shear flow 
{\it Physical Review E} 61: 6621-6631.

\bibitem{erzan} Derrida B, Eckmann JP, Erzan A
(1983) Renormalisation groups with periodic and aperiodic orbits. 
{\it Journal of Physics A} 16: 893-906 (1983).

\bibitem{Gunsdiam} Diamond JM (1997) {\it Guns, germs, and 
steel : the fates of human societies.} W.W. Norton \& Co. New York.

\bibitem{DombGreen}  Domb C, Green M.S. (1976)
{\it Phase transitions and critical phenomena.} Academic Press, London, 
New York.

\bibitem{Drozdz} Drozdz S, Ruf F, Speth J 
Wojcik M (1999) Imprints of log-periodic self-similarity in the stock market.
{\it European Physics Journal B} 10: 589-593.

\bibitem{Dubrulle} Dubrulle B, Graner F,
Sornette D (1997) {\it Scale invariance and beyond.} EDP Sciences and 
Springer, Berlin.

\bibitem{Elias} Elias D (1999) {\it Dow 40,000 : Strategies for 
Profiting from the Greatest BullMarket in History.} McGraw-Hill ?

\bibitem{comment} The fits have been performed using 
the ``amoeba-search''
algorithm (see  {\it Numerical Recipes} by  W.H. Press, B.P. Flannery, S.A.
Teukolsky and W.T. Vetterling, Cambridge University Press, Cambridge
UK, 1992) minimizing the variance of the fit to the data. We stress that all
three linear variables $A$, $B$ and $C$ are slaved to the other nonlinear
variables
by imposing the condition that, at a local minimum, the variance has zero first
derivative with respect these variables. Hence, they should not be regarded as
free parameters, but are calculated solving three linear equations using
standard techniques including pivoting. Note in addition that the phase $\phi$
in (\ref{lppow}) is just a (time) unit as are the coefficients $A$, $B$ and
$C$. The key physical variables are thus $t_c$, $\beta$ and $\omega$.

\bibitem{vonFoerster} von Foerster 
H, Mora P.M, Amiot L. W (1961) Population Density and Growth. {\it Science}
133: 1931-1937

\bibitem{Doomsdaypaper} von 
Foerster H, Mora P.M, Amiot L.W (1960) Doomsday: Friday 13 November
A.D. 2026. {\it Science} 132: 1291-1295.

\bibitem{cycles} More information 
about the foundation can be found at http://www.cycles.org/cycles.htm. However,
it seems that the foundation is not very active presently.

\bibitem{Glassman} Glassman JK, Hassett KA (1999)
{\it DOW 36,000: The New Strategy for Profiting from the Coming Rise in the 
Stock Market.} Times Books ?


\bibitem{global} Global Financial Data, Freemont 
Villas, Los Angeles, CA 90042. The data use are free samples available at
http://www.globalfindata.com/.

\bibitem{critical} Hahne F.J (1983) {\it Critical Phenomena, 
Lecture Notes in Physics 186} page 209. Springer, Berlin, Heidelberg.

\bibitem{Hanson} Hanson R (2000) Could it happen again? 
Long-term growth as a sequence of exponential modes. Working paper 
available at http://hanson.gmu.edu/longgrow.html.

\bibitem{Hern} Hern W.M (1993) Is human culture carcinogenic for 
uncontrolled population growth and ecological destruction? 
{\it BioScience} 43: 768-773. He concludes that the sum of human activities, 
viewed over the past tens of thousand of years, exhibits all four major 
characteristics of a malignant process: rapid uncontrolled growth; invasion 
and destruction of adjacent tissues (ecosystems, in this case); metastasis 
(colonization and urbanization, in this case); and dedifferentiation (loss
of distinctiveness in individual components as well as communities throughout 
the planet).

\bibitem{Herroux} Herrmann, H.J. and Roux, S., editors (1990)
Statistical models for the fracture of disordered media (Amsterdam; 
New York: North-Holland ; New York, N.Y., U.S.A.)

\bibitem{failure} Johansen A, Sornette D (1998)
Evidence of discrete scale invariance by canonical averaging. {\it 
International Journal of Modern Physics C} 9: 433-447 and references therein.

\bibitem{bubmodel} Johansen A, Sornette D (1999)
Critical crashes. {\it Risk Magazine} 12: 91-94.

\bibitem{crash} Johansen A, Sornette D
Ledoit 0 (1999) Predicting Financial Crashes using discrete scale invariance.
{Journal of Risk} 1: 5-32 and references therein.

\bibitem{JLS}  Johansen A, Ledoit O, 
Sornette (2000) Crashes as critical points. {\it International Journal of 
Theoretical and Applied Finance} 3: 219-255.

\bibitem{faicri} Johansen A, Sornette D (2000)
Critical ruptures. {\it European Physics Journal B} 18: 163-181
(e-print at http://arXiv.org/abs/cond-mat/0003478)

\bibitem{earthquake} Johansen A, Saleur H, 
Sornette D (2000) New Evidence of Earthquake Precursory Phenomena in the  17 
Jan. 1995 Kobe Earthquake, Japan. {\it European Physics Journal B} 15: 551-555
and references therein.

\bibitem{emerg} Johansen A, Sornette D (2000) 
Log-periodic power law bubbles in Latin-American and Asian markets and 
correlated anti-bubbles in Western stock markets: An empirical study. 
in press in Int. J. Theor. Appl. Finance. Available at 
http://arXiv.org/abs/cond-mat/9907270 

\bibitem{nasdaq} Johansen A, Sornette D (2000)
The Nasdaq crash of April 2000: Yet another example of log-periodicity in a 
speculative bubble ending in a crash. {\it European Physics Journal 
B} 17,: 319-328
(e-print at http://arXiv.org/abs/cond-mat/0004263).

\bibitem{Kadlec} Kadlec CW (1999) {\it Dow 100,000: Fact or 
Fiction} Prentice Hall Press ?.

\bibitem{Kapitza} Kapitza SP (1996), Phenomenological theory 
of world population growth. {\it Uspekhi Fizichskikh Nauk} 166: 63-80.

\bibitem{Kremer} Kremer M (1993) Population growth and 
technological change: One million B.C. to 1990. {\it Quarterly Journal of 
Economics} 108: 681-716.

\bibitem{Mehra} Mehra R, Prescott E (1985) Title ?
{\it Journal of Monetary Economics 15}: 145-161.

\bibitem{Moffatt} Moffatt H.K (2000) Euler's disk and its 
finite-time singularity. {\it Nature} 404: 833-834.

\bibitem{Pumiersiggia} Pumir A,  Siggia E.D (1992) 
Vortex morphology and Kelvin theorem. {\it Physical Review A} 45: R5351-5354.

\bibitem{Raan} van Raan AFJ (2000) On growth, ageing and 
fractal differentitation of science. {\it Scientometrics} 47: 347-362.

\bibitem{Rascle} Rascle M, Ziti C (1995) Finite-time 
blow-up in some models of chemotaxis. {Journal of Mathematical Biology} 33: 
388-414.

\bibitem{Ritz} Rietz TA, Mehra R,
 Prescott EC (1988) The Equity Risk Premium: A Solution?
{\it Journal of Monetary Economics} 22: 117-136.

\bibitem{Romer} Romer D(1996) {\it Advanced macroeconomics.}
McGraw-Hill Companies New York.

\bibitem{Romer2} Romer PM (1990) Endogeneous technological change.
{\it Journal of Political Economy} 98: S71-S102.

\bibitem{salsor} Saleur, Sornette (1996)
Complex exponents and log-periodic corrections in frustrated systems.
{\it Journal de Physique I France} 6: 327-355.

\bibitem{SSS} Saleur H, Sammis CG, Sornette D
(1996) Renormalization group theory of earthquakes. {\it Nonlinear Processes
in Geophysics} 3: 102-109.

\bibitem{statementAcads} 
Science Summit on World Population: A Joint Statement by 58 of the 
World's Scientific Academies (1994) {\it Population and Development Review}
20: 233-238.

\bibitem{Simon} Simon J.L (1996) {\it The Ultimate Resource 2?} 
Princeton University Press, Princeton, NJ.

\bibitem{SJ97} Sornette D, Johansen A (1997) Large 
financial crashes. {\it Physica A} 245: 411-422.

\bibitem{reviewsor} Sornette D (1998) Discrete scale 
invariance and complex dimensions. {\it Physics Reports} 297: 239-270.

\bibitem{Nego} {\it Earth negotiation 
Bulletin} Vol. 15, No. 34 March 27 2000. Available at:
http://www.iisd.ca/linkages/download/pdf/enb1534e.pdf.

\bibitem{UShistory} Even though U.S.A was recognised as a nation by 
the Paris Treaty in 1783, a number of events point to the fact it was not 
{\it established} as a nation before $\approx$  1790. They are as follows. 
1) The constitution went into effect in March 1789, having been ratified by 
New Hampshire as the ninth state on June 21, 1788. 2) The last of the thirteen
states, Rhode Island, first approved it on May 29, 1790. 3) The first census 
in the U.S. was made in 1790. 4) The Naturalisation Act of 1790 grants the 
right of U.S. citizenship to all ``free white persons.'' 5) In 1790, the
Federal Government declared that it was redeeming the SCRIP MONEY that was
issued during the Revolutionary War. 6) At about this time, the Government
announced the creation of the first bank of the United States in conjunction 
with the sale of \$10,000,000 dollars in shares of stock.

\bibitem{Chaline2} Nottale L., Chaline J., Grou P. (2000) Les arbres de l'\'evolution
(Hachette Litterature, Paris) 379 p.

\bibitem{Chaline3} Nottale L., Chaline J., Grou P. (2000) in ``Fractals 2000 in Biology
and Medicine'', Proceedings of Third International Symposium, Ascona,
Switzerland, March 8-11, 2000, Ed. G. Losa, BirkhŠuser Verlag.

\bibitem{Meyer1} Meyer F. (1947) L'acc\'el\'eration \'evolutive.
Essai sur le rythme \'evolutif et son 
interpr\'etation quantique. Librairie des Sciences et des Arts, Paris, 67p.

\bibitem{Meyer2} [14] Meyer F. (1954) Probl\'ematique de l'\'evolution. P.U.F., 279p.

\bibitem{Grou} Grou P. (1987,1995) L'aventure \'economique. L'Harmattan, Paris, 160 p.

\bibitem{Braudel} Braudel F. (1979) Civilisation mat\'erielle, \'economie et capitalisme. 
A. Colin 

\bibitem{Gilles} Gilles B. (1982) Histoire des techniques. Gallimard


\end{thebibliography}
\end{document}